\documentclass[12pt,a4paper]{scrbook}
\usepackage[dvips]{graphicx}
\begin{document}
\normalsize
\def \i {\mathrm{i}}

\title {The Conformally Invariant Vector Field on  $R \times S^3$}
\author{ Zurab Ratiani\\
\\
Diploma Thesis at the \\
Institut f\"ur Theoretische Physik \\
der Universit\"at Hannover \\}
\date{May  2002}
\maketitle

\tableofcontents

\chapter{Introduction}

Conformal symmetry  is the highest space-time  symmetry. In addition to the transformations of the Poincare group it includes the transformations which preserve angles between  world lines:dilatations and the special conformal transformations.The last transformation is singular in the  Minkowski space, but on   $R \times S^3 $ it is well defined. Global space-time may differ from Minkowski space, while being locally very similar to it, and the new  ''energy'', the translation operator $P_0$ from the conformal algebra, will  correspondingly differ from the special relativistic energy, although again  being very close to it in the case of sufficiently localized phenomena. The difference in energy-eigenvalues should be very small for localized states, but there is a considerable lost of symmetry in the replacement of   $R \times S^3 $  by Minkowski space. \\

The second chapter deals with   geometrical objects, with  curvature related quantities: Riemann tensor, Ricci tensor, Ricci scalar.   The conformal Killing-equations are presented.
 Next, the electromagnetic field will be constructed on the manifold $R \times S^3$. This manifold is a special case of the Robertson-Walker spaces. $R \times S^3$ has  $15$ generators of the conformal group and therefore  is a maximal conformally symmetric manifold. The space of the rays on the lightcone in  $R^{2,4}$  is isomorphic  to the $( S \times  S^3)/ Z_2$. This fact is helpful to find out the conformal Killing-vectors, that describe the conformal symmetry of the curved manifold.  $R \times S^3 $ is a universal covering space of  $( S \times  S^3)/ Z_2$.

In the third chapter the electromagnetic field is considered. The Lagrangian   that is invariant under the conformal transformations of the dynamical fields $A_m$  only is written. The Maxwell field in four dimensions is a vector field and in contrast to the scalar field there is no need to introduce  densities in order to obtain a conformally invariant  theory. For the eigenmode expansion of the field  $A_m$ the eigenfunctions of the ''four-momentum''  operator, the operator that plays the role  of the momentum in the conformal algebra, will be used.
At the end of this chapter, a gauge-fixing term is found, which makes strongly apparent the similarity to the flat case.

In the fourth chapter the Green's function is found. It allows us to carry out the quantization. Here  the creation and annihilation operators of   photons on the curved space will be discussed. The interaction of the Maxwell field  with the charged scalar  density fields $\phi(x)$ is considered. The question whether such terms  break   the conformal invariance,  or give raise to  some new vertex-factors for the Feynman diagrams is investigated. Finally  the energy-spectrum  of the dynamical field $A_m(x)$ will be  considered. Due to  the gauge invariance  of the  field  there exist   the unphysical degrees of freedom which lead to the negative-norm states.  The BRS  method will  be  used  to solve this problem.

\chapter{Mathematical background}

\section{The Conformal transformations}
Suppose we have a region $M$  of the space $R^n$ with coordinates $x_1,  \dots , x_n$,  and some mapping $T$ of points $x$ to  points $x'$ : $x'=T(x)$. On $M$   tangential vectors to  curves are defined :
\begin{eqnarray}
\frac{dx'^m}{ds }_{\mid_ {x'(x)}}=\frac{\partial x'^m}{\partial x^n} \frac{d x^n}{ds}_{ \mid_ x}
\end{eqnarray}

An invertible transformation is called conformal if the length-square of   all vectors will change at most by  some coordinate  dependent factor $e^{\lambda(x)}$:
\begin{eqnarray}
\frac{\partial x'^m}{\partial x^k} \frac{d x^k}{ds}_{ \mid_ x} \frac{\partial x'^n}{\partial x^l} \frac{d x^l}{ds}_{ \mid_ x} g_{mn}(x'(x))= e^{\lambda(x)}  \frac{d x^k}{ds}_{ \mid_ x}\frac{d x^l}{ds}_{ \mid_ x} g_{kl}(x)
\end{eqnarray}
This must be valid for every  vector, so:
\begin{eqnarray}
\label{kli1}
\frac{\partial x'^m}{\partial x^k}  \frac{\partial x'^n}{\partial x^l} g_{mn}(x'(x))=e^{\lambda(x)} g_{kl}(x)
\end{eqnarray}

The last formula shows us that after a conformal transformation the new metric tensor $g'_{mn}(x')$  is proportional to the original  one  $g_{mn}(x') $ if both are  expressed in the same coordinates.
The transformation is called isometry if the conformal factor $e^{\lambda(x)}=1$ and hence length-square of the vector does not change.
 Equation (\ref{kli1}) describes the finite conformal transformations.\\

The one parameter-group of transformations $T_{\alpha}$ maps   points  $x$ to   points  $ x'$  of the manifold. If one varies the  group parameter $\alpha$,  one gets a  curve with the tangential vectors:
\begin{eqnarray}
\xi^m(T_{\alpha}(x))=\frac{(T_{\alpha}x)^m}{d \alpha}.
\end{eqnarray}

 Inserting $\alpha=0$  one obtains  vector at the original point $x$.

 Considering  infinitesimally small transformations:  $x'^m=x^m+\xi^m$, if one differentiates (\ref{kli1}) with respect to  the transformations parameter at  $\alpha=0$ and uses the equations:
\begin{eqnarray}
\frac{\partial (T_{\alpha}x)^m}{\partial x^k}_{\mid_{\alpha=0}}=\delta^m_k, \qquad \partial_{\alpha} \frac{\partial (T_{\alpha}x)^m}{\partial x^k}_{\mid_{\alpha=0}}=\partial_k \xi^m
\end{eqnarray}
one gets:
\begin{eqnarray}
\partial_k  \xi^m \delta^n_l g_{mn}+\delta_k^m \partial_l \xi^n g_{mn}+\delta_k^m \delta_l^n \partial_{\alpha}g_{mn}(x)+\epsilon g_{kl}=0.
\end{eqnarray}
Here we have used the  notation: $\epsilon=-\partial_{\alpha} \lambda_{\mid_{\alpha=0}}$.

Finally  one receives the conformal Killing equation  :
\begin{eqnarray}
\label{kli2}
\xi^k \partial_k g_{mn}+\partial_m \xi^k g_{kn}+\partial_n \xi^k g_{mk}+\epsilon g_{mn}=0.
\end{eqnarray}

From the last equations one can find the  conformal Killing vectors, if the components of the metric tensor are known.  This vectors describe the symmetry of the manifold.  Because of the symmetry of the metric tensor (\ref{kli2}) there are  at most $d(d+1)/2$ equations, where $d $ is the dimension of the manifold.

\section{Isometries}
Isometries are transformations that  do not change  lengths  of  vectors.  The  Killing vectors generate such transformations.  This vectors can be found from the Killing equation.  The last one   is  received from the equation (\ref{kli2}) if one sets  the parameter $\epsilon=0$:
\begin{eqnarray}
\label{kil}
\xi^k \partial_k g_{mn}+\partial_m \xi^k g_{kn}+\partial_n \xi^k g_{mk}=0.
\end{eqnarray}

If the  components of the metric tensor are known  one can find the Killing vectors from equations (\ref{kil}).  We will compute the components of  one Killing vector.  In our case, for manifold $R \times S^3 $ we get the following equations:
\begin{eqnarray}
& & (11) \qquad \partial_1 \xi^1=0  \\
  \nonumber\\
& &  (22) \qquad    \xi^1  \cos \alpha \partial_2 \xi^2 \sin \alpha=0
\end{eqnarray}

\begin{eqnarray}
\label{33}
 (33) \qquad  \xi^1 \cos \alpha \sin \theta+\xi^2 \sin \alpha \cos \theta\partial_3+\xi^3 \sin \alpha \sin \theta=0
\end{eqnarray}

\begin{eqnarray}
(12) \qquad \partial_1 \xi^2 \sin^2 \alpha +\partial_2 \xi^1=0
\end{eqnarray}

\begin{eqnarray}
\label{13}
 (13)  \qquad \partial_1\xi^3 \sin^2 \alpha \sin^2 \theta +\partial_3 \xi^1=0
\end{eqnarray}

\begin{eqnarray}
\label{23}
(23)  \qquad \partial_2\xi^3 \sin^2 \alpha \sin^2 \theta+\partial_3 \xi^2 \sin^2 \alpha =0.
\end{eqnarray}

The numbers in  brackets indicate the indices of the metric tensor.  Here we assume   that the numbers $0, 1, 2, 3 $ correspond to the coordinates $t, \alpha, \theta ,\phi $ accordingly.
 To solve this system of equations we will use the  Ansatz:
\begin{eqnarray}
& & \xi^1=A\sin (\theta+a)  \nonumber\\
& & \xi^2=A\cos (\theta+a) \cot \alpha +b
\end{eqnarray}
Here $A,a,b$  are some unknown constants.  From the equation (\ref{33}) one gets:
\begin{eqnarray}
\xi^3=-\phi  \Bigl(A\sin (\theta+a) \cot \alpha +(A\cos (\theta+a) \cot \alpha+b) \cot  \theta  \Bigr).
\end{eqnarray}
If we insert this into  (\ref{13}) we will receive the following condition onto the  a's:  $ a=\frac{ \pi n}{2}$,  where $ n= 0, \pm 1, \pm 2,\pm 3, \dots $.
From  equation (\ref{23}) we receive  for  $a=\pi/2$  that  $ b=0$.  Now  we can insert this values for the constants $a$ and  $b$   into  the last  expression for  $\xi^3$ and we get  $\xi^3=0$.  So one  can now write down the following Killing vector:
\begin{eqnarray}
\xi^m=(\cos \theta,-\sin \theta \cot \alpha,0).
\end{eqnarray}

\section{Covariant Differentiation}

The gradient of a tensor does not behave like a tensor under general coordinate transformations $x^i=x^i(y^1, \dots y^n), i=1 \dots n$.  The   components of a tensor $\tilde T_k$ in the $y^i$ coordinates satisfy:
\begin{eqnarray}
\tilde T^j_{,l}&=&\frac{\partial \tilde T_k}{\partial y^l }=\frac{\partial }{\partial y^l } \left( T^i  \frac{\partial x^j }{\partial y^i}  \right)=\frac{\partial T^i }{\partial y^l }  \frac{\partial y^j }{\partial x^i}+T^i \frac{\partial  }{\partial y^l} \frac{\partial y^j }{\partial x^i }  \nonumber \\
& &=\frac{\partial T^i }{\partial x^p }  \frac{\partial x^p }{\partial y^l} \frac{\partial y^j }{\partial x^i}+T^i \frac{\partial^2 y^j}{\partial x^i  \partial x^k } \frac{\partial x^k}{\partial y^l }.
\end{eqnarray}

The second term  vanishes under linear coordinate transformations  $y^i=a^i_j x^j$, but not under   general coordinate  transformations.  So,  it is necessary to generalize the  concept of the gradient.
The covariant derivative of (1,1) tensor has the form:
\begin{eqnarray}
D_k T^i_j=\frac{\partial T^i_j }{\partial x^k}+\Gamma^i_{lk} T^i_j-\Gamma^l_{jk} T^i_l,
\end{eqnarray}
and there exist similar expressions  for tensors of  different rank.
 Covariant differentiation is a linear operation and commutes with the operation of contraction.  For a zero-rank tensor it is the  usual gradient of the function.  The covariant derivative of a product of tensors is calculated using the  Leibniz product rule.

The $\Gamma$-s are called   Christoffel symbols or affine connections.  The coordinates are said to be (Psuedo)Euclidean (with respect to a given connection), if in terms of those coordinates  the $\Gamma$'s are identically zero:
\begin{eqnarray}
\label{eu1}
\Gamma^i_{kl} \equiv 0  \qquad i,j,k=1, \dots ,n.
\end{eqnarray}

On the other hand, the coordinates $x^1,\dots ,x^n$ are said to be Euclidean if in terms of these coordinates the metric $g_{ij}$ is given by:
\begin{eqnarray}
\label{eu2}
 g_{ij}=\delta_{ij},
\end{eqnarray}
where $\delta $  is the Kronecker symbol.

A connection $\Gamma^i_{jk}$ is said  to be compatible with a metric $ g_{ij}$  if the covariant derivative of the metric tensor  $ g_{ij}$ is identically zero:
\begin{eqnarray}
D_k g_{ij} \equiv 0  \qquad i,j,k=1, \dots ,n.
\end{eqnarray}

For a vector field :
\begin{eqnarray}
D_kD_lT^i&=& D_k \left(\frac{\partial T^i}{\partial x^l}+\Gamma^i_{nl}T^n \right)=\frac{\partial}{\partial x^k} \left(\frac{\partial T^i}{\partial x^l}+\Gamma^i_{nl}T^n \right) +\Gamma^i_{pk} \left(\frac{\partial T^p}{\partial x^l}+\Gamma^p_{nl}T^n \right)- \nonumber \\
& & -\Gamma^p_{lk} \left(\frac{\partial T^i}{\partial x^p}+\Gamma^i_{np}T^n \right)  \nonumber \\
&=&\frac{\partial^2 T^i}{\partial x^k  \partial x^l}+\frac{\partial T^n}{\partial x^k} \Gamma^i_{nl}+ \Gamma^i_{pk} \frac{\partial T^p}{\partial x^l}-\Gamma^p_{lk} \frac{\partial T^i}{\partial x^p}+T^n \frac{\partial \Gamma^i_{nl}}{\partial x^k}+  \nonumber \\
& &   \nonumber \\
& &+\Gamma^i_{pk} \Gamma^p_{nl}T^n - \Gamma^p_{lk} \Gamma^i_{np}T^n.
\end{eqnarray}

If one exchanges the  order of differentiation and after that subtracts this two expressions from each other, one gets:
\begin{eqnarray}
(D_kD_l-D_lD_k)T^i&=& \left( \frac{\partial \Gamma^i_{nl}}{\partial x^k}-\frac{\partial \Gamma^i_{nk}}{\partial x^l} \right) T^n+\left(\Gamma^i_{pk} \Gamma^p_{nl}-\Gamma^i_{pl} \Gamma^p_{nk} \right) T^n \nonumber \\
& &-(\Gamma^p_{lk}-\Gamma^p_{kl}) \left( \frac{\partial T^i}{\partial x^p}+\Gamma^i_{np}T^n  \right).
\end{eqnarray}
 Introducing the notation:
\begin{eqnarray}
\label{curvature}
-R^i_{nkl}&=& \frac{\partial \Gamma^i_{nl}}{\partial x^k}-\frac{\partial \Gamma^i_{nk}}{\partial x^l} +\Gamma^i_{pk} \Gamma^p_{nl}-\Gamma^i_{pl} \Gamma^p_{nk}, \\
T^p_{kl}&=& \Gamma^p_{lk}-\Gamma^p_{kl},
\end{eqnarray}

 finally one obtains :
\begin{eqnarray}
\label{covcomm}
(D_kD_l-D_lD_k)T^i=-R^i_{nkl} T^n+T^p_{kl} D_n T^i
\end{eqnarray}

One can write   analogous equations for the tensors with  lower indices:
\begin{eqnarray}
\label{covcom}
(D_kD_l-D_lD_k)T_i=R^n_{ikl} T_n+T^p_{kl} D_n T_i.
\end{eqnarray}

The $R^i_{nkl}$ are the components of so-called  Riemann curvature tensor and  $T^p_{kl}$ is the torsion tensor.  If the connection is symmetric  $T^p_{kl} \equiv 0 $.

One can write the conformal Killing equation (\ref{kli2}) in the  following form:
\begin{eqnarray}
D_m \xi_n +D_n \xi_m+ \epsilon g_{mn}=0,
\end{eqnarray}
where the $\xi$'s are the conformal Killing vectors.  If one multiplies  both sides with the inverse metric tensor   one gets: $\epsilon =-\frac{2}{d} D_k \xi^k$, where $d$ is the  dimension of the space.

For isometries one obtains  the Killing equation:
\begin{eqnarray}
\label{covis}
D_m \xi_n +D_n \xi_m=0.
\end{eqnarray}

If one uses  (\ref{covis})  in the equation for the commutator of the covariant derivatives (\ref{covcom})  one gets: $ D_k D_l \xi_i+D_l D_i \xi_k=-R^n_{ikl} \xi_n $.  (We assume that $ T\equiv  0 $).  Using  the Bianchi-identity $D_{(k} D_l \xi_{i)}=0$  one receives:

\begin{eqnarray}
D_i D_k \xi_l=-R^n_{ikl} \xi_n
\end{eqnarray}

So the second derivatives of the Killing vectors are completely determined  through this  vectors and the  Riemann tensor.  In the Taylor expansion of the analytical Killing vector fields  around a point P  at most the coefficients $\xi$ and the antisymmetric coefficients $D_k \xi_{l \mid p}  $  are free. In $d$ space-time dimensions these  are  in total  $d+ \frac{1}{2}d(d-1)$ coefficients and hence  there exist  at most $\frac{1}{2}d(d+1)$  linear independent Killing vectors.


 \section{The manifold $R \times S^3$}

As $R \times S^3$ we will consider the following overlapping of the manifold $S^1 \times S^3$ which  is a submanifold of the flat space $R^{2,4}$:
\begin{eqnarray}
\label{embed}
& &(t, \alpha ,\theta ,\phi) \rightarrow (y^{-1},y^0,y^1,y^2,y^3,y^4 )=  \nonumber \\
& &(\cos t,\sin t,  \sin \alpha \sin \theta \cos \phi,\ sin \alpha \sin \theta  \sin \phi , \sin \alpha \cos \theta , \cos \alpha  )
\end{eqnarray}
Here $y$'s are the  coordinates on  $R^{2,4}$ and  $S^1 \times S^3$ is described through the equation:
\begin{eqnarray}
\sum_{i=-1}^{0}  (y^i)^2 =\sum_{i=1}^{4}  (y^i)^2=1
\end{eqnarray}

The angles $\alpha $ and $\theta$ take  values from $0$ to $ \pi$,  $\phi$  from $0$  to $2 \pi$, and   $t$ takes all real numbers.  From $R^{2,4}$  one can write down  the metric on  $R \times S^3$ :
\begin{eqnarray}
& &\left(\frac{dy}{ds} \frac{dy}{ds}\right)=\left(\frac{dy^{-1}}{ds}\right)^2+\left(\frac{dy^0}{ds}\right)^2-\left(\frac{dy^1}{ds}\right)^2-\left(\frac{dy^2}{ds}\right)^2-\left(\frac{dy^3}{ds}\right)^2-\left(\frac{dy^4}{ds}\right)^2  \nonumber \\
& &=\left(\frac{dt}{ds}\right)^2 -\left(\frac{d \alpha}{ds}\right)^2 -\sin^2 \alpha\left(\frac{d \theta}{ds}\right)^2-\sin^2 \alpha   \sin^2 \theta  \left(\frac{d \phi}{ds}\right)^2
\end{eqnarray}
hence :
\begin {displaymath}
g_{mn}=
\left( \begin{array}{cccc}
1  &    0  &    0  &    0    \\
0  &    -1  &   0  &    0  \\
0  &     0  &     - \sin^2 \alpha & 0 \\
0  & 0  &0  & - \sin^2 \alpha   \sin^2 \theta
\end{array}  \right).
\end {displaymath}

\section{Curvature of  $R \times S^3$}
The curvature of a manifold is described through the Riemann tensor:
\begin{eqnarray}
\label{rim2}
R^l_{ijk}=\frac{\partial \Gamma^l_{ij}}{\partial x^k}-\frac{\partial \Gamma^l_{ik}}{\partial x^j}+\Gamma^m_{ij} \Gamma^l_{mk}-\Gamma^m_{ik} \Gamma^l_{ml}
\end{eqnarray}
which  are called Riemann symbols of the second kind,
where  the Christoffel symbols $\Gamma$'s are:

\begin{eqnarray}
\label{gam}
\Gamma^m_{kl}=\frac{1}{2}g^{mn}(\partial_k g_{nl}+\partial_l g_{nk}-\partial_n g_{kl})
\end{eqnarray}

The components of the corresponding covariant tensor of  fourth order :
\begin{eqnarray}
R_{mjik}=g_{mh} R^h_{jik}
\end{eqnarray}
are called the Riemann symbols of the first kind. They satisfy the following identities:
\begin{eqnarray}
\label{risy}
& &R_{mjik}=-R_{jmik}, \\
& &R_{mjik}=-R_{mjki}, \\
& &R_{mjik}=R_{ikmj},
\end{eqnarray}
and
\begin{eqnarray}
\label{bia}
R_{mjik}+R_{mikj}+R_{mkji}=0.
\end{eqnarray}
From (\ref{risy})  it follows that for non-vanishing components neither  the first two nor the second two indices can be  alike. Due to  (\ref{risy}) there are $n(n-1)/2 \equiv n_2$ cases  in which the first pair of indices are equal to the second pair, and $ n_2(n_2-1)/2$ cases in which the first pair and the second pair  are different.  Hence there are total $ n_2(n_2+1)/2$  distinct symbols as regards  (\ref{risy}).  Additionally there are $n(n-1)(n-2)(n-3)/4! \equiv n_4 $  equations of the form (\ref{bia}).  Consequently there are
\begin{eqnarray}
\label{rnum}
N=n_2(n_2+1)/2-n_4=n^2(n^2-1)/12
\end{eqnarray}
distinct symbols of the first kind.

For the space $R \times S^3$ one  gets that the componets of the Christoffel symbols are:
\begin{eqnarray*}
& &\Gamma^{\phi}_{\alpha \phi}=\Gamma^{\phi}_{\phi \alpha }=\cot \alpha \\
& &\Gamma^{\theta}_{\alpha \theta}=\Gamma^{\theta}_{\theta \alpha}=\cot \alpha  \\
& &\Gamma^{\phi}_{\theta \phi}=\Gamma^{\phi}_{\phi \theta}=\cot \theta  \\
& &\Gamma^{\alpha}_{\theta \theta}=-\sin \alpha \cos \alpha  \\
& &\Gamma^{\alpha}_{\phi \phi}=-\sin \alpha \cos \alpha  \sin^2 \theta  \\
& &\Gamma^{\theta}_{\phi \phi}=- \sin \theta  \cos \theta
\end{eqnarray*}
All other components vanish.

The number of the non-vanishing Riemann symbols of the first kind are less than predicted by the formula (\ref{rnum}) because of the symmetries of our space. They have the following form:
\begin{eqnarray}
& &R_{\alpha,\theta,\alpha,\theta}=-\sin^2 \alpha  \nonumber \\
& & R_{\alpha,\phi,\alpha,\phi}=-\sin^2 \alpha  -\sin^2 \theta  \nonumber \\
& &R_{\theta,\phi,\theta,\phi}=-\sin^4 \alpha  -\sin^2 \theta  \nonumber
\end{eqnarray}
and other components  that  can be received  through exchange of the indices, in accordance  with (\ref{risy}), which makes  $12$ components in total.

The Ricci tensor is computed through the formula:
\begin{eqnarray}
 R_{mn}=g^{kl} R_{kmln}.
\end{eqnarray}

For $R \times S^3 $  the Ricci tensor takes the form:
\begin{eqnarray}
R_{mn}=
\left( \begin{array}{cccc}
0  &    0  &    0  &    0    \\
0  &    -2  &   0  &    0  \\
0  &     0  &     -2 \sin^2 \alpha & 0 \\
0  & 0  &0  & -2 \sin^2 \alpha   \sin^2 \theta
\end{array}  \right).
\end{eqnarray}

The Ricci Scalar is:
\begin{eqnarray}
R=g^{mn}R_{mn}
\end{eqnarray}
and for $R \times S^3 $ one gets   $  R=6 $.

One  can compute   the curvature of  $R \times S^3$ in a different way.  The curvature is completely determined through its three dimensional curvature tensor $P_{plmn}$  of  $S^3$.  The manifold  $S^3$   belongs to the class of the maximally  symmetric spaces.  The Riemann tensor of such a completely isotropic space is:
\begin{eqnarray}
P_{plmn}=k(g_{pm}g_{ln}-g_{lm}g_{pn}),
\end{eqnarray}
where $ k$ is some constant.
Then the Ricci tensor is:
\begin{eqnarray*}
P_{mn}=2k g_{mn},
\end{eqnarray*}
and  the Ricci scalar:
\begin{eqnarray}
P=6k.
\end{eqnarray}

The line element on $S^3$ is: $dl^2= \gamma_{mn}dx^m dx^n$, where $\gamma_{mn}$ is the  three-dimensional metric. So, the curvature   of such spaces are described only through the constant $ k$, and we will try to find its value for  $S^3$.  One can consider $S^3$ as embedded in some four-dimensional flat space. The equation of the hyper-sphere with unit radius has the form:
\begin{eqnarray}
\label{sph}
x_1^2+ x_2^2+ x_3^2+x_4^2=1,
\end{eqnarray}
and the line element on it is:  $dl^2=dx_1^2+ dx_2^2 + dx_3^2+dx_ 4^2$.
With the help of (\ref{sph}) one can rearrange the last formula to give:
\begin{eqnarray}
dl^2=dx_1^2+ dx_2^2 + dx_3^2+\frac{(x_1dx_1+x_2 dx_2+x_3dx_3)^2 }{1-x_1^2- x_2^2- x_3^2},
\end{eqnarray}
so the metric tensor is:
\begin{eqnarray}
\gamma^{mn}=\delta^{mn}+\frac{x^m x^n}{1-x_1^2- x_2^2- x_3^2}.
\end{eqnarray}
One can choose to compute the curvature of  $S^3$ at the origin, this means at the north pole.  The curvature at any other point of the sphere is the same. So in the neighborhood  of the north pole the first derivatives of the metric $\gamma_{mn}$ vanish and so do  the quantities $\gamma_{mn}$. The second derivatives are:
\begin{eqnarray}
\partial_m \partial_n \gamma_{kl_{\mid x=0}}=\delta_{mk}\delta_{nl}+\delta_{ml}\delta_{nk},
\end{eqnarray}
 and the Ricci tensor is:
\begin{eqnarray}
R_{ik_{\mid x=0}}=\partial_l\Gamma^l_{ik}- \partial_k\Gamma^l_{il}.
\end{eqnarray}
Using (\ref{gam}) one gets:
\begin{eqnarray*}
R_{ik}&=&1/2 \gamma^{lm}(\partial_l \partial_i \gamma_{mk}+\partial_l \partial_k \gamma_{mi}-\partial_l \partial_m \gamma _{ik}) \nonumber \\
& &-1/2 \gamma^{lm}(\partial_k \partial_i \gamma_{ml}+\partial_k \partial_l \gamma _{mi}-\partial_k \partial_m \gamma_{il}) \nonumber \\
&= & \gamma^{lm} \delta_{lm}\delta_{ik} - \gamma^{lm} \delta_{km} \delta_{il}=2 \delta_{ik}.
\end{eqnarray*}
Finally the scalar curvature  is: $R=R_{ik} \gamma^{ik}=6. $

\section{Generators of the conformal transformations}

We can now  write down the  Generators of $SO(2,4)$:
\begin{eqnarray}
L_{-1,0}&=& \partial_t \nonumber  \\
L_{-1,1}&=& - \sin t \sin \alpha \sin \theta \cos \phi \partial_t+\cos t \cos \alpha \sin \theta \cos \phi \partial_\alpha  \nonumber  \\
 & &\frac{\cos t\cos \theta \cos \phi}{\sin \alpha} \partial_\theta-\frac{\cos t \sin \phi}{\sin \alpha \sin \theta}\partial_\phi  \nonumber  \\
L_{-1,2}&=& - \sin t \sin \alpha \sin \theta \sin \phi \partial_t+\cos t \cos \alpha \sin \theta \sin \phi \partial_\alpha  \nonumber  \\
 & &\frac{\cos t\cos \theta \sin \phi}{\sin \alpha} \partial_\theta+\frac{\cos t \cos \phi}{\sin \alpha \sin \theta}\partial_\phi  \nonumber  \\
L_{-1,3}&=&-\sin t \sin \alpha \cos \theta \partial_t+\cos t \cos \alpha \cos \theta \partial_\alpha-\frac{\cos t \sin \theta}{\sin \alpha} \partial_\theta    \nonumber  \\
L_{-1,4}&=&-\sin t \cos \alpha \partial_t-\cos t \sin \alpha \partial _\alpha   \nonumber  \\
L_{0,1}&=& \cos t \sin \alpha \sin \theta \cos \phi \partial_t+\sin t \cos \alpha \sin \theta \cos \phi \partial_\alpha  \nonumber  \\
 & &\frac{\sin t\cos \theta \cos \phi}{\sin \alpha} \partial_\theta-\frac{\sin t \sin \phi}{\sin \alpha \sin \theta}\partial_\phi  \nonumber  \\
L_{0,2}&=& \cos t \sin \alpha \sin \theta \sin \phi \partial_t+\sin t \cos \alpha \sin \theta \sin \phi \partial_\alpha  \nonumber  \\
 & &\frac{\sin t\cos \theta \sin \phi}{\sin \alpha} \partial_\theta+\frac{\sin t \cos \phi}{\sin \alpha \sin \theta}\partial_\phi  \nonumber  \\
L_{0,3}&=& \cos t \sin \alpha \cos \theta \partial_t+\sin t \cos \alpha \cos \theta \partial_\alpha-\frac{\sin t \sin \theta}{\sin \alpha} \partial_\theta    \nonumber  \\
L_{0,4}&=& \cos t \cos \alpha \partial_t - \sin t \sin \alpha \partial_\alpha    \nonumber  \\
L_{1,2}&=&  \partial_\phi  \nonumber  \\
L_{2,3}&=& \sin \phi \partial_\theta + \frac{\cos \theta \cos \phi}{\sin \theta}  \partial_\phi \nonumber  \\
L_{3,1}&=& -\cos \phi \partial_\theta + \frac{\cos \theta \sin \phi}{\sin \theta} \partial_\phi \nonumber  \\
L_{1,4}&=&\sin \theta \cos \phi \partial_\alpha+\cot \alpha \cos \theta \cos \phi \partial_\theta-\cot \alpha \frac{\sin \phi}{\sin \theta}\partial_\phi   \nonumber  \\
L_{2,4}&=&\sin \theta \sin \phi \partial_\alpha+\cot \alpha \cos \theta \sin \phi \partial_\theta+\cot \alpha \frac{\cos \phi}{\sin \theta}\partial_\phi   \nonumber  \\
L_{3,4}&=& \cos \theta \partial_\alpha-\cot \alpha \sin \theta \partial_\theta   \\ & &  \nonumber  \\
 L_{a,b}&=&-L_{b,a}
\end{eqnarray}

In Appendix A, as an example, $L_{0,3}$  is computed. \\

The generators D of dilatations, K of special conformal transformations,  and $  P_m,J_{mn} $ of the Poincar\'e group fulfill the following commutation relations:
\begin{eqnarray}
\label{cal}
& &[P_m,P_n]=0  \hspace{3cm}  [J_{mn},P_k]=-(\eta_{mk}P_n-g_{nk}P_m) \nonumber  \\
& & [J_{mn},J_{kp}]=\eta_{mp}J_{nk}+\eta_{nk}J_{mp}-\eta_{mk}J_{np}-\eta_{np}J_{mk} \nonumber  \\
& &[D,P_m]=-P_m  \hspace{3cm} [D,J_{mn}]=0   \nonumber  \\
& &[D,K_m]=K_m  \hspace{3.2cm}   [K_n,K_m]=0     \nonumber \\
& &[K_m,P_n]=-2(\eta_{mn} D+J_{mn}) \hspace{1cm} [J_{mn},K_p]=-(\eta_{mp}K_n-\eta_{np}K_m)
\end{eqnarray}
where  $\eta_{mn}$ is the Minkowski metric.
One can verify that this algebra is fulfilled through the following combinations of  the Killing vectors of $SO(2,4)$:
\begin{eqnarray}
\label{gen}
P_m&=&L_{-1,m}+L_{m,4},  \hspace{1cm}  K_m=L_{-1,m}-L_{m,4},  \nonumber  \\
 J_{mn}&=&L_{m,n}, \hspace{3cm} D=-L_{-1,4}
\end{eqnarray}

In appendix B one of this commutator relations   is proven.


\section {Noether's  Theorem}

The   physical  systems can have  a symmetry under  transformations of the dynamical fields, this means that under corresponding infinitesimal  transformations $ \delta \phi_i$ the action of the system does not change: $\delta W=0$.   If  $ T_{\alpha}$  is a one-parameter group that maps  fields into  fields,  then  infinitesimal  transformations are defined as the derivatives of the transformations at $\alpha=0 $:
\begin{eqnarray}
\delta \phi_i=\frac{d}{d\alpha}T_{\alpha}\phi_i(x)_{\mid_{\alpha=0}}.
\end{eqnarray}

The action will change as:
\begin{eqnarray}
\delta W=\int d^4x \delta \phi_i \frac{\delta W}{\delta \phi_i}.
\end{eqnarray}
The transformation  $ T_{\alpha}$ is called a symmetry of the action if the integrand in the last expression can be written as the total derivative of some function:
\begin{eqnarray}
\label{symm}
\delta \phi_i \frac{\delta W}{\delta \phi_i}+\partial_m j^m=0
\end{eqnarray}

To the current the   term  $j^m=\partial_n B^{nm}$ ,  the total divergence of the functions that  are antisymmetric under the exchange of the indices,  can be added.  For  physical fields  the equations  of  motion  are $\frac{\delta W}{\delta \phi_i}=0$,  so to each symmetry there corresponds a conserved current $j^m$.  The variation of the Lagrangian is:
\begin{eqnarray}
\delta  \mathcal{L}&=&\delta \phi_l \frac{\partial \mathcal{L}}{\partial \phi_l}+ (\partial_k \delta \phi_l) \frac{\partial \mathcal{L}}{\partial (\partial_k \phi_l)} \nonumber \\
&=& \delta \phi_l \left(\frac{\partial \mathcal{L}}{\partial \phi_l}-\partial_k \frac{\partial \mathcal{L}}{\partial (\partial_k \phi_l)}\right)+\partial_k \left( \delta \phi_l \frac{\partial \mathcal{L}}{\partial (\partial_k \phi_l )} \right).
\end{eqnarray}

The term in  brackets  in the first term is called the Euler derivative:
\begin{eqnarray}
\frac{\hat \partial \mathcal{L}}{\hat \partial \phi_l}=\frac{\partial \mathcal{L}}{\partial \phi_l}-\partial_k \frac{\partial \mathcal{L}}{\partial (\partial_k \phi_l)},
\end{eqnarray}
so,  the variation of the Lagrangian is:
\begin{eqnarray}
\delta  \mathcal{L}&=&\delta \phi_l \frac{\hat \partial \mathcal{L}}{\hat \partial \phi_l}+\partial_k \left( \delta \phi_l \frac{\partial \mathcal{L}}{\partial (\partial_k \phi_l )} \right) \nonumber \\
&=&\partial_m K^m ,
\end{eqnarray}
where:
\begin{eqnarray}
K^m= \delta \phi_l \frac{\partial \mathcal{L}}{\partial (\partial_m \phi_l )}-j^m.
\end{eqnarray}
and  (\ref{symm}) was used. From here  the formula for  the conserved current follows:
\begin{eqnarray}
\label{concurrent}
j^m=\delta \phi_l \frac{\partial \mathcal{L}}{\partial (\partial_m \phi_l )}-K^m+\partial_n B^{nm}.
\end{eqnarray}
A transformation of the fields is a symmetry transformation if the corresponding variation of the Lagrangian is the total derivative of some function.
 On the other hand to each conserved currents there corresponds a symmetry transformation of the action. One can write:
\begin{eqnarray}
0=\partial_m j^m+R_l \frac{\delta W}{\delta \phi_l}+R_l^m \partial_m \left( \frac{\delta W}{\delta \phi_l} \right),
\end{eqnarray}
where $R_l,R_l^m$  are some functions of the coordinates, the fields and their derivatives. One can bring the last equation into the form:
\begin{eqnarray}
0=\partial_m \left( j^m+R_l^m  \frac{\delta W}{\delta \phi_l}  \right) +  \left( R_l-\partial_m R_l^m \right) \frac{\delta W}{\delta \phi_l},
\end{eqnarray}
according to (\ref{symm}) to the conserved current there corresponds a symmetry transformation:
\begin{eqnarray}
\delta \phi_l=R_l-\partial_m R_l^m.
\end{eqnarray}
The corresponding conserved charge is:
\begin{eqnarray}
Q(t)= \int d^3x j^0(t,\vec x).
\end{eqnarray}

\section{The flat coordinates }

The Killing-vectors corresponding to the momentum operators of the conformal algebra (\ref{gen}) are:
\begin{eqnarray}
\label{veckill}
\xi^n_0 & = & \Big[ 1+\cos  t  \cos  \alpha  ,  \quad -\sin  t  \sin  \alpha , \quad 0,  \quad 0 \Bigr] \nonumber \\
\xi^n_1 & = & \Big[ -\sin t  \sin  \alpha  \sin  \theta  \cos  \phi , \quad  \sin  \theta  \cos
  \phi   ( 1+\cos  t  \cos
\alpha )  , \nonumber \\
 & & {\frac {\cos  \theta  \cos
  \phi   ( \cos  t  +\cos  \alpha)
    }{\sin  \alpha  }}, \quad  -{\frac {\sin
  \phi   ( \cos  t  +\cos  \alpha)
    }{\sin  \alpha  \sin  \theta
  }}\Bigr] \nonumber\\
\xi^n_2 & = & \Big[ -\sin  t  \sin  \alpha  \sin
\theta  \sin  \phi , \quad \sin  \theta
\sin  \phi  (  1+\cos  t  \cos
\alpha )  ,\nonumber\\
 & &  \quad {\frac {\cos  \theta  \sin
  \phi  (  \cos  t  +\cos  \alpha)
    }{\sin  \alpha  }}, {\frac {\cos
\phi   ( \cos  t  +\cos  \alpha)
    }{\sin  \alpha  \sin  \theta
  }}\Bigr] \nonumber \\
\xi^n_3 & = & \Big[ -\sin  \alpha  \sin  t  \cos  \theta ,\quad \cos  \theta(1+  \cos  t  \cos  \alpha   ) ,  \nonumber\\
& &  -{\frac {\sin  \theta
  (  \cos  t  +\cos  \alpha )
  }{\sin  \alpha  }},  0\Bigr]
\end{eqnarray}

Here we introduce the following quantities:
\begin{eqnarray}
\beta^m&=&(t,\alpha,\theta,\phi),
\end{eqnarray}
\begin{eqnarray}
\label{xcoord}
x^m&=&\Biggl( \frac{\sin t}{\cos t +\cos \alpha},\frac{\sin \alpha \sin \theta \cos \phi}{\cos t +\cos \alpha},\frac{\sin \alpha \sin \theta \sin \phi}{\cos t +\cos \alpha}, \frac{\sin \alpha \cos \theta}{ \cos t +\cos \alpha}\Biggr)  \\
\nonumber \\
 & & m=0,1,2,3.  \nonumber
\end{eqnarray}
In terms of this expression we can write the Killing vectors as:
\begin{eqnarray}
\xi^n_a=\frac {\partial\beta^n}{\partial x^a}.
\end{eqnarray}

Here we define the  inverse Killing-vectors in the following sense:
\begin{eqnarray}
& &\bar \xi_m^a=\frac{\partial x^a}{\partial \beta^m}, \\
& &\bar \xi_m^a \xi_b^m=\delta_b^a , \hspace{2cm}  \bar \xi_m^a \xi_a^n=\delta_m^n.
\end{eqnarray}

So, one has four inverse  vectors corresponding to the  vectors in (\ref{veckill}):

\begin{eqnarray}
\label{invkil}
\bar \xi^0_n & = & \Big[ \frac{1+\cos  t  \cos  \alpha }{(\cos t+\cos \alpha)^2} , \quad \frac{\sin  t  \sin  \alpha}{(\cos t+\cos \alpha)^2} , \quad 0,  \quad 0 \Bigr]     \nonumber     \\
\bar \xi^1_n & = & \Big[ \frac{\sin t  \sin  \alpha  \sin  \theta  \cos  \phi}{(\cos t+\cos \alpha)^2} , \quad  \frac{\sin  \theta  \cos
  \phi   ( 1+\cos  t  \cos
\alpha )}{(\cos t+\cos \alpha)^2}  , \nonumber \\
 & & \frac {\sin \alpha \cos  \theta  \cos
  \phi    }{\cos t+\cos \alpha)  }, \quad  -\frac {\sin \alpha \sin \theta \sin \phi  }{\cos t+ \cos \alpha  }\Bigr]  \nonumber \\
\bar \xi^2_n & = &\Big[    \frac{\sin t  \sin  \alpha  \sin  \theta  \sin  \phi}{(\cos t+\cos \alpha)^2} , \quad  \frac{\sin  \theta  \sin
  \phi   ( 1+\cos  t  \cos
\alpha )}{(\cos t+\cos \alpha)^2}  , \nonumber \\
 & & \frac {\sin \alpha \cos  \theta  \sin
  \phi    }{\cos t+\cos \alpha  }, \quad  \frac {\sin \alpha \sin \theta \cos  \phi  }{\cos t+ \cos \alpha  }     \Bigr]        \nonumber \\
\bar \xi^3_n & = & \Big[ \frac{\sin t  \sin  \alpha  \cos  \theta  }{(\cos t+\cos \alpha)^2} , \quad  \frac{\cos \theta    ( 1+\cos  t  \cos
\alpha )}{(\cos t+\cos \alpha)^2}  , \nonumber \\
 & & -\frac {\sin \alpha \sin \theta   }{\cos t+\cos \alpha  }, \quad  0 \Bigr]\end{eqnarray}

In our case the following relations hold:
\begin{eqnarray}
\label{mtavari}
& &g_{mn}\xi^m_a \xi^n_b=(\cos t+\cos \alpha)^2 \eta_{ab}, \\
& &g^{mn} \bar \xi_m^a \bar \xi_n^b=(\cos t+\cos \alpha)^{-2} \eta^{ab}, \\
& &\bar \xi^a_m=\xi^n_b g_{mn} \eta^{ab}(\cos t+\cos \alpha)^{-2}.
\end{eqnarray}

The Jacobian of the transformation from flat  $x^m$ coordinates to  curved coordinates $ \beta^m $ is:
\begin{eqnarray}
\label{jacobian}
det_4 \parallel \partial x/ \partial \beta \parallel & = & \frac{\sin^2 \alpha \sin \theta}{(\cos \theta +\cos \alpha)^4},  \\
det_3 \parallel \partial x/ \partial \beta \parallel & = & det_4 \parallel \partial x/ \partial \beta \parallel (1+\cos t \cos \alpha).
\end{eqnarray}

In Minkowski space the line element has the form:
\begin{eqnarray}
ds^2=(dx^0)^2-(dx^1)^2-(dx^2)^2-(dx^3)^2
\end{eqnarray}
 One can rewrite this in the spherical polar coordinates $(t,r,\theta,\phi)$:
\begin{eqnarray}
ds^2=dt^2-dr^2-r^2(d \theta^2+\sin^2 \theta d \phi^2 )
\end{eqnarray}

This metric is singular for $r=0$  and $\sin \theta=0$ but this is only a  coordinate singularity.  To get rid of it one  has to restrict the coordinates to the ranges: $0<r< \infty,  0<\theta< \pi, 0<\phi<2 \pi $ and one needs two such coordinate patches to cover the whole Minkowski space.
 Further one introduces the advanced and retarded null coordinates:
\begin{eqnarray}
\label{advret}
& &v=t+r  \nonumber \\
& &w=t-r
\end{eqnarray}
Then one has:
\begin{eqnarray}
ds^2=dvdw -\frac{1}{4}(v-w)^2 (d \theta^2+\sin^2 \theta d \phi^2 )
\end{eqnarray}
 Also  this coordinates can take  infinite values.  In the new coordinates $p$ and $q$:
\begin{eqnarray}
& &\tan p=v  \nonumber \\
& & \tan q=w  \nonumber \\
& &-\frac{1}{2} \pi<p<\frac{1}{2} \pi, \quad -\frac{1}{2} \pi<q<\frac{1}{2} \pi , \quad p \ge q
\end{eqnarray}
this infinities are transformed to finite values and the metric takes the form:
\begin{eqnarray}
ds^2=\sec^2 p \sec^2 q(dpdq-\frac{1}{4} \sin^2(p-q) (d \theta^2+\sin^2 \theta d \phi^2 ))
\end{eqnarray}

One defines  new coordinates:
\begin{eqnarray*}
t'=p+q, \quad r'=p-q
\end{eqnarray*}
with
\begin{eqnarray}
\label{borders}
-\pi<t'+r'< \pi, \quad -\pi<t'-r'< \pi, \quad  r' \ge 0
\end{eqnarray}

The original metric $g$ is conformal to the metric $\bar g $ given by:
\begin{eqnarray}
\label{einstein}
d \bar s^2=(dt')^2-(dr')^2- \sin^2 r' (d \theta^2+\sin^2 \theta d \phi^2 ))
\end{eqnarray}
Thus the  Minkowski space-time is given by the region (\ref{borders}) of the metric:
\begin{eqnarray}
d  s^2=\frac{1}{4} \sec^2 (t'+r') \sec^2 (t'-r') d \bar s^2
\end{eqnarray}


One can see that the metric (\ref{einstein}) is locally identical to that of the space $R \times S^3 $,
 Einstein  static universe. In the later the $r'$ coordinate is not limited. Suppressing two of the coordinates
 one can represent  Einstein's  static universe as the cylinder $x^2+y^2=0$
\newpage
 in the three-dimensional Minkowski space: $ds^2=-dt^2+dx^2+dy^2$.




\section{Infinitesimal transformations}

Under coordinate transformation the components of  tensors of the second rank transform as:
\begin{eqnarray}
\label{finit}
F_m^{'n}(y)=\frac{\partial x^k}{\partial y^m} \frac{\partial y^n}{\partial x^l}F_k^l(x).
\end{eqnarray}

For an infinitesimal transformation:
\begin{eqnarray}
y^m=x^m+\xi^m,
\end{eqnarray}
(\ref{finit}) in the first order of $\xi^m$ takes the form:
\begin{eqnarray}
\delta_{\xi} F_k^l=\xi^m \partial_m F_k^l(x)+\partial_k \xi^m F_m^l(x)-\partial_m \xi^l F_k^m(x).
\end{eqnarray}
This expression is called the Lie derivative along a vector field $\xi^m$. For  tensors of different rank the derivative has an analogous form.

The tensor densities transform in different way under the coordinat transformations:
\begin{eqnarray}
D_m^{'n}(y)=\parallel\frac{\partial x}{\partial y}\parallel^{n/4} \frac{\partial x^k}{\partial y^m} \frac{\partial y^n}{\partial x^l}D_k^l(x).
\end{eqnarray}

The quantity $n$ in the power of the Jacobian is the conformal weight of the tensor density. The infinitesimal transformations for such objects is of  the form:
\begin{eqnarray}
\delta_{\xi} D_k^l=\xi^m \partial_m d_k^l(x)+\partial_k \xi^m D_m^l(x)-\partial_m \xi^l D_k^m(x)+\frac{n}{4} \partial_m \xi^m D_k^l.
\end{eqnarray}

The conformal Killing equatio can be shortly written as:
\begin{eqnarray}
\delta_{\xi} g_{mn}=\epsilon  g_{mn}.
\end{eqnarray}


\chapter{Classical field theory}

\section{The vector Field}

The Lagrangian for an  electromagnetic field is:
\begin{eqnarray}
\label{lagmax}
\mathcal{L}_{\rm{vec}}=-\frac{1}{4}\sqrt{g}g^{mk}g^{nl}F_{mn}F_{kl},
\end{eqnarray}
where:
\begin{eqnarray}
F_{mn}=\partial_m A_n-\partial_n A_m.
\end{eqnarray}

The equations of motion are:
\label{max1}
\begin{eqnarray}
\partial_m(\sqrt{g}  F^{mn})=0.
\end{eqnarray}

This Lagrangian is invariant under gauge transformations of the type:
\begin{eqnarray}
A_m \rightarrow A_m+\partial_m \lambda.
\end{eqnarray}

The form of the Lagrangian shows us that the $ A_m$ are vectors (no densities) and they  transform    under the general coordinate transformations   according to:
\begin{eqnarray}
\label{DELTAA}
\delta A_m= A_m^{\prime}(x)-A_m(x)=\xi^l \partial_l A_m+\partial_m \xi^l A_l.
\end{eqnarray}
Due to  the gauge-freedom we may add  a  divergence term to it:
\begin{eqnarray}
\label{delta_a}
\delta^{\prime} A_m&=&\delta A_m- \partial_m(\xi^l A_l)=\xi^l F_{lm}     \\
\Rightarrow  \delta^{\prime } F_{mn}&=&\partial_m(\xi^l F_{ln})-\partial_n(\xi^l F_{lm} ).
\end{eqnarray}

We will  now prove that the  action is invariant under (\ref{DELTAA}) for  $\xi$'s corresponding to   conformal transformations.  The Lagrangian of the electro-magnetic field transforms as a scalar density of  weight $4$ under general coordinate transformations, so the variation:
\begin{eqnarray}
\label{deltal}
\delta_G \mathcal{L}&=&\partial_m(\xi^m \mathcal{L})=\partial_m(\xi^m \sqrt{g}F_{mn}F^{mn}),
\end{eqnarray}
and so the variation  of the action vanishes: $\delta_G W=0$.

Under   conformal transformations  the metric tensor  transforms as $\delta g_{mn} =\epsilon g_{mn}$.  One can check that the expression $ g^{1/4} g^{mk}  $ is invariant under   conformal transformations:
\begin{eqnarray}
\delta ( g^{1/4}g^{mk})&=&\delta ( g^{1/4}) g^{mk}+  g^{1/4} \delta g^{mk} \nonumber \\
&=&\frac{1}{4}g^{1/4} g^{nl} \epsilon g_{nl} g^{mk}-g^{1/4} \epsilon g^{mk}=0
\end{eqnarray}
because $g^{nl} g_{nl}=4$.  From this one gets:
\begin{eqnarray*}
\delta (\sqrt{g}g^{mk} g^{np})=\delta ( g^{1/4} g^{mk})  g^{1/4} g^{np}+ g^{1/4} g^{mk} \delta(  g^{1/4} g^{np}) \nonumber \\
&=&0
\end{eqnarray*}

So one can conclude that  conformal transformations of the whole  action  $(3.1) $ are  equal to   conformal transformations   of the dynamical fields $A_m$ only.  Because of this and (\ref{deltal}) the action is invariant under  conformal transformations of the fields   $A_m$. \\

The equation of motion (\ref{max1}) can be written with covariant derivatives:
\begin{eqnarray}
\partial_m(\sqrt{g}  F^{mn})=\sqrt{g}(\partial_m F^{mn}+\Gamma_{ml}^m F^{ln}+\Gamma_{ml}^nF^{ml})=\sqrt{g}D_m F^{mn},
\end{eqnarray}
here is used that the partial derivative: $\partial_m \sqrt{g}=\frac{1}{2 \sqrt{g}} g g^{kl} \partial_m g_{kl}=\sqrt{g} \, \Gamma_{mk}^k $ and the fact that an  antisymmetric tensor times a symmetric tensor vanishes,  so one can add the last term.
Hence  the Maxwell equation is:
\begin{eqnarray}
\label{max2}
\sqrt{g}D_m F^{mn}=0.
\end{eqnarray}

Since the  Christoffel symbols $\Gamma's$ are symmetrical in the lower indices one can rewrite the stress tensor $F_{mn}$ with  covariant derivatives: $F_{mn}=D_m A_n-D_n A_m$ and the equations of motion take the form:
\begin{eqnarray}
D^m (D_m A_n-D_n A_m)=0.
\end{eqnarray}
We would like to rearrange the second term. The covariant derivatives do not commute. After  formula (\ref{covcomm}) with vanishing torsion one has:
\begin{eqnarray}
D_m D_n A^m=[D_m,D_n]A^m+D_n D_m A^m=R_{mnr}^m A^r+D^n D_M A^m.
\end{eqnarray}
In the second term the trace of the Riemann tensor  gives the Ricci tensor, so the equations of motion are:
\begin{eqnarray}
D^m D_m A_n+R_{nm}A^m-D_n D_m A^m=0.
\end{eqnarray}
One can see that in curved space the Maxwell equation contains a term with the Ricci tensor that had no analogous term in the flat space equations.

\bigskip
We can express the fields $A$ as the superposition of the momentum eigenfunctions $ u_m$, which are defined by the following equation:
\begin{eqnarray}
\label{eigvalue}
\xi^l_a \partial_l u_m+\partial_m \xi^l_a u_l=-\i k_a u_m,
\end{eqnarray}
where the  $\xi^l_a  $ are  the  Killing-vectors  corresponding to  the momentum generator  $ P_a $ in the conformal algebra. We will now show that this equation is fulfilled by:
\begin{eqnarray}
u_m=\partial_m (\varepsilon x) \exp(-\i kx),
\end{eqnarray}
where   x's  are the $ \beta $  dependent functions we have introduced above. \\
Proof:
\begin{eqnarray}
& &\xi^l_a \partial_l u_m+\partial_m \xi^l_a u_l \nonumber \\
&=& \frac{\partial}{\partial x^a}u_m+\partial_m(\xi^n_a u_n)-\xi^n_a \partial_m u_n   \nonumber \\
&=&  \frac{\partial u_m}{\partial x^a}+ \partial_m ( \varepsilon_b \xi^n_a \bar \xi^b_n \exp(-\i kx))-\xi^n_a (\partial_m \partial_n \varepsilon x)\exp(-\i kx)-\nonumber \\
& &-\xi^n_a\partial_n (\varepsilon x) \partial_m \exp{(-\i kx)} \nonumber \\
&=& \frac{\partial u_m}{\partial x^a}+ \varepsilon_a \partial_m \exp{(-\i kx)}-\xi^n_a \partial_n \Bigl( \partial_m (\varepsilon x) \exp{(-\i kx)}\Bigr) +\nonumber \\
& &+\xi^n_a\partial_m (\varepsilon x)\partial_n \exp{(-\i kx)}- \varepsilon_a \partial_m \exp{(-\i kx)} \nonumber \\
&=& \frac{\partial u_m}{\partial x^a}-\frac{\partial u_m}{\partial x^a}+\partial_m(\varepsilon x)\frac{\partial}{\partial x^a}\exp{(-\i kx)} \nonumber \\
&=& \partial_m(\varepsilon x)(-\i k_a) \exp{(-\i kx)}   \nonumber \\
&=& -\i k_a u_m .   \nonumber   \\
& &  \hspace{7cm} \nonumber
\end{eqnarray}

Here we have used that through  the inverse Killing-vectors the solution can be rewritten in the form:
\begin{eqnarray}
u_m=\varepsilon_a \bar \xi^a_m  \exp{(-\i kx)}.
\end{eqnarray}

\section{The gauge-fixing term}

One  defines the  tensor $ f_{mn}  $ as:
\begin{eqnarray}
f_{mn}&=&\partial_m u_n-\partial_n u_m   \nonumber   \\
&=& \i  \varepsilon_a k_b(\bar \xi^a_m  \bar \xi^b_n - \bar \xi^a_n \bar \xi^b_m ) \exp(-\i kx ).
\end{eqnarray}
Inserting this terms into the Maxwell equations one  gets:
\begin{eqnarray}
\label{pirveli}
D_m f^{mn}&=& D_m \left(g^{km} g^{ln} \i  \varepsilon_a k_b (\bar \xi^a_k  \bar \xi^b_l - \bar \xi^a_l \bar \xi^b_k ) e^{-\i kx}  \right)   \nonumber   \\
&=& g^{km} g^{ln} \i  \varepsilon_a k_b  D_m(\bar \xi^a_k  \bar \xi^b_l - \bar \xi^a_l \bar \xi^b_k ) e^{-\i kx}  )+  g^{km} g^{ln} \i  \varepsilon_a k_b (\bar \xi^a_k  \bar \xi^b_l - \bar \xi^a_l \bar \xi^b_k ) D_m  e^{-\i kx}  )    \nonumber   \\
&=& \Bigl(\i g^{km} g^{ln} \varepsilon_a k_b D_m(\bar \xi^a_k  \bar \xi^b_l - \bar \xi^a_l \bar \xi^b_k  ) -\frac{\eta^{cb} }{(\cos t +\cos \alpha)^2}k_b k_c \varepsilon_a  \bar \xi^a_l g^{ln}+ \nonumber   \\
& &+ \frac{\eta^{ca} }{(\cos t +\cos \alpha)^2}\varepsilon_a k_c k_b  \bar \xi^b_l g^{ln} \Bigr) e^{-\i kx}.
\end{eqnarray}
It can be shown that   first term is zero.    We see that the second term is proportional to $k^2  $  and  the last term  to $\varepsilon k  $ .

Setting  (\ref{pirveli}) equal to zero we obtain  conditions for the k's and $\varepsilon$'s such that the u's are  special solutions of the equations of motion.  \\
 We would like to impose a gauge-condition upon the u's that leads to $\varepsilon k=0 $.  In order to find such a condition,  we consider the following condition for some unknown function $f(\beta)$:
\begin{eqnarray}
\label{meore}
\partial_m (f u^m)&=&\partial_m \left(f \varepsilon_a \bar \xi^a_n g^{mn} e^{-\i kx} \right)  \nonumber   \\
&=& -\i k_b \varepsilon_a f g^{mn} \bar \xi^a_n \bar \xi^b_m  e^{-\i kx}+\partial_m( f g^{mn}  \bar \xi^a_n)\varepsilon_a e^{-\i kx}  \nonumber   \\
&=& -\i k_b \varepsilon_a  \eta^{ab} \frac{1}{(\cos t +\cos \alpha)^2} f e^{-\i kx}+ \nonumber   \\
& &+\partial_m( f g^{mn} \bar \xi^a_n)\varepsilon_a e^{-\i kx}\nonumber   \\
&=&0.
\end{eqnarray}
As we see the first term is proportional to $ \varepsilon k $.  We will now find an $ f(\beta) $  such that the second term vanishes. This leads to the following:
\begin{eqnarray}
\label{fcomput}
\partial_m( f g^{mn} \bar  \xi^a_n)=g^{mn} \bar \xi^a_n \partial_m f+f\partial_m \bar  \xi^{am}     &=&0 \nonumber   \\
 g^{mn} \bar \xi^a_n \partial_m f&=&-f \partial_m(\bar  \xi^{am}) \nonumber   \\
\xi^l_a  g^{mn}\bar \xi^a_n \partial_m f&=&-\xi^l_a f \partial_m(\bar  \xi^{am}) \nonumber   \\
  g^{lm} \partial_m f&=&-f \xi^l_a  \partial_m(\bar \xi^{am}) \nonumber   \\
\partial_p f&=&-f  \xi_{ap}  \partial_m( \bar \xi^{am}).
\end{eqnarray}
This equation is satisfied by :
\begin{eqnarray}
f=\frac{\sqrt{g}}{(\cos t+\cos \alpha)^2}.
\end{eqnarray}
So we choose the gauge-condition to be :
\begin{eqnarray}
\partial_m (\frac{\sqrt{g}}{(\cos t+\cos \alpha)^2}  A^m)=0.
\end{eqnarray}
From equation  (\ref{pirveli}) we get a  condition for $k_0  $:
\begin{eqnarray}
k_0=\sqrt{k_1^2+k_2^2+k_3^2}
\end{eqnarray}
The $\varepsilon_a $'s can be developed in basis of linearly independent vectors $\varepsilon_a^v $.

Then the A's can be written as :
\begin{eqnarray}
\label{A_m}
A_m=\int \frac{d^3k}{(2 \pi)^3 \sqrt{2k_0}}( a_v \varepsilon_a^v \bar \xi^a_m \exp(-\i kx)+  a_v^+  \varepsilon_a^ { v} \bar \xi^a_m \exp(\i kx) ).
\end{eqnarray}

We now enlarge the Lagrangian by a gauge-fixing term:

\begin{eqnarray}
\label{fix}
\mathcal{L}=-\frac{1}{4}F_{mn}F^{mn}- \frac{1}{2\sqrt{g}T^2}[\partial_m(\sqrt{g}TA^m)]^2,
\end{eqnarray}

where:
\begin{eqnarray}
\label{tdef}
T=\frac{1}{(\cos t+\cos \alpha)^2},
\end{eqnarray}

The equations of motion now take the form:
\begin{eqnarray}
\label{equmo}
D_m(F^{mn})+ T g^{mn}\partial_m(\frac{1}{\sqrt{g} T^2}\partial_l(\sqrt{g} T A^l))=0.
\end{eqnarray}

The gauge-fixing term  is a scalar density of the weight $4$,  so it is invariant under general coordinate transformation $\delta _G$.  One can rewrite the gauge-fixing term in the form:
\begin{eqnarray}
\mathcal{L}_{\rm{gauge}}=\frac{1}{(g^{1/4}T)^2}[\partial_m(\sqrt{g} g^{mn} T A_n)]^2.
\end{eqnarray}

The term $g^{1/4} T$ has  weight  $2$ and one can show that for   Poincar\'e transformations:
\begin{eqnarray}
\delta_P (g^{1/4} T)=\xi^l \partial_l (g^{1/4} T)+\frac{1}{2}\partial_l \xi^l (g^{1/4} T)=0,
\end{eqnarray}
 with the Killing-vectors corresponding to the translation and the Lorentz transformations from the conformal algebra (\ref{gen}).
Using  the last equality, the fact that for the conformal transformations $\delta_{c} (g^{1/4} g^{mn})=0$ and the chain rule, finally one gets:
\begin{eqnarray}
0=\delta_P \mathcal{L}_{\rm{gauge}}= \frac{1}{(g^{1/4}T)^2}[\partial_m(\sqrt{g} g^{mn} T \delta_P A_n)]^2
\end{eqnarray}
So the  gauge-fixing term is invariant under the Poincar\'e transformations. However it is not invariant under conformal transformations.

\chapter{Quantum  field theory}

\section{ Propagator quantization}

The Green's  function of the equation of motion (\ref{equmo})  is:
\begin{eqnarray}
\label{green}
G_{mn}(x,x')&=&\int \frac{d^4 k}{(2 \pi )^4} \frac{-\i }{k^2+ \i \varepsilon} u_m^a(x) u^{b \ast}_n \eta_{ab} \nonumber \\
&=& \int \frac{d^4 k}{(2 \pi )^4} \frac{-\i }{k^2+ \i \varepsilon} \bar  \xi^a_m(x) \bar  \xi^b_n(x') \eta_{ab} \exp{(\i k(x-x'))},
\end{eqnarray}
with $ u^a_m \equiv \bar \xi^a_m \exp(-\i kx)$.  \\
To prove this  we insert  $G_{mn}$ in the equation of motion (\ref{equmo}).  The $x$ dependent part under  the integral is $ u^a_m \eta_{ab}$.  In the following computation we  will use only this part. With the results of the computations in (\ref{pirveli}) and (\ref{meore}), replacing $\epsilon_a $ by $\eta_{ab}$, one gets:
\begin{eqnarray}
& &D_m \left(g^{sm} g^{ln}(\partial_s u^a_l \eta_{ab}-\partial_l u^a_s \eta_{ab} \right) + Tg^{mn} \partial_m \left(\frac{1}{\sqrt{g}T^2}\partial_l (\sqrt{g}T g^{ls} u_s^a \eta_{ab})  \right)=  \nonumber \\
&=&-k^2 T \eta_{ab} \bar \xi^a_l g^{ln} e^{-\i kx}+ \eta^{ca} \eta_{ab} T k_c k_d \bar \xi^d_l g^{ln} e^{-\i kx}- \nonumber \\
& &-Tg^{mn} \partial_m \left(\frac{1}{\sqrt{g}T^2} \i k_d \eta_{ab} \eta^{ad} T^2 \sqrt{g} e^{-\i kx} \right) \nonumber \\
&=&-k^2 T \eta_{ab} \bar \xi^a_l g^{ln} e^{-\i kx}.
\end{eqnarray}
  Finally the action of the operator of the equation of motion $L^{ln}$ on the Green's function gives:
\begin{eqnarray}
L^{ln} G_{lr}= \int \frac{d^4 k}{(2 \pi )^4} \frac{\i }{k^2+ \i \varepsilon}  \bar  \xi^b_r(x')  k^2 T \eta_{ab} \bar \xi^a_l(x) g^{ln}  \exp{(\i k(x-x'))}=\i \delta^n_r  \delta^4(x-x').
\end{eqnarray}
Here according to  (\ref{mtavari}) it has been used that: $T \eta_{ab} \bar \xi^a_l(x) \bar  \xi^b_r(x)=g_{lr} $.

One can perform the $k_0$ integration in the Green's function (\ref{green}):
\begin{eqnarray}
\label{Gmn}
G_{mn}(x)=-\int \frac{d^3 k}{(2 \pi )^3 2 k_0} \eta_{ab}  \bar  \xi^a_m(x) \bar  \xi^b_n(x') \left\{\theta(x^0-x'^0) e^{-\i k(x-x')}+\theta (x'^0-x^0)  e^{\i k(x-x')} \right\}
\end{eqnarray}
with $k_0=\sqrt{k_1^2+k_2^2+k_3^2}$.

On the other side one has the expression:
\begin{eqnarray}
& &G_{mn}(x)=\langle 0 \vert \theta(x^0-x'^0)A_m(x)A_n(x')+\theta(x'^0-x^0) A_n(x')A_m(x)  \vert 0  \rangle   \nonumber \\
&= &\int \frac{d^3 k}{(2 \pi )^3 \sqrt{2 k_0}} \int \frac{d^3 p}{(2 \pi )^3 \sqrt{2 p_0}}  \Bigl\{ \theta(x^0-x'^0)  [a_a(k),a_b^+(p)] \bar \xi^a_m(x) \bar \xi^b_n(x')  e^{-\i kx+ \i px' }+  \nonumber \\
& & \hspace{2cm} +\theta(x'^0-x^0)  [a_a(p),a_b^+(k)] \bar \xi^a_n(x') \bar \xi^b_m(x)  e^{\i kx-\i px'}  \Bigr\}.
\end{eqnarray}

The comparison with the formula (\ref{Gmn}) shows that the creation and annihilation operators must fulfill the equality:
\begin{eqnarray}
\label{acomutat}
[a_a(k),a_b^+(p)]= -\eta_{ab} \delta^3(k-p)
\end{eqnarray}

From  one side the momentum operator acts on the field as:
\begin{eqnarray}
[\i P_b,A_m]=\int \tilde k   [\i P_b,a_a] \bar \xi^a_m e^{-\i kx}+ [\i P_b,a^+_a] \bar \xi^a_m e^{\i kx}
\end{eqnarray}

On the other  side we know how the momentum  operator acts on  its eigenmodes $u(x)$:
\begin{eqnarray}
\delta_b A_m=\int \tilde k  a_a   (-\i k_b) \bar \xi^a_m e^{-\i kx}+ a^+_a   \i k_b \bar \xi^a_m e^{\i kx}
\end{eqnarray}

Comparing the last two formulas one gets for the operator:
\begin{eqnarray}
[\i P_b,a^+_a]=\i k_b a_a^+
\end{eqnarray}
showing that $a^+_a$ is indeed a creation operator, in the sense that it creates momentum.


\section {Creation and annihilation operators}

In this section $a$ and $a^+$'s will be expressed as functions of the fields and its canonical momentum. For mathematical simplicity computations will be carry out for  $ x_0=0$.
 The mode expansion of the field for the polarisazion vector: $\varepsilon_a^{\nu}=\delta_a^{\nu}$ is:
\begin{eqnarray}
\label{AA_m}
A_m=\int \frac{d^3k}{(2 \pi)^3 \sqrt{2k_0}}( a_a(k) \bar \xi^a_m e^{-\i kx}+   a^+_a(k) \bar \xi^a_m e^{\i kx} ).
\end{eqnarray}

Multiplying both sides with a term and integrating over the $\beta=(t,\alpha,\theta,\phi)$ curved coordinates one gets:
\begin{eqnarray*}
& &\int d^3 \beta \sqrt{g} T \bar \xi^0_0 \bar \xi_b^m e^{\i px} A_m = \nonumber \\
&=&\int  \frac{d^3k}{(2 \pi)^6 \sqrt{2k_0}} \int d^3 \beta  \sqrt{g} T \bar \xi^0_0 ( a_b(k)  e^{-\i(k-p)x}+   a^+_b(k)  e^{\i (k+p)x}),
\end{eqnarray*}
where  $T$ is given  through the formula (\ref{tdef}) and  the zero component of the  inverse Killing-vector is:  $ \bar \xi^0_0=(1-\cos t \cos \alpha)/(\cos t+ \cos \alpha)^2$.   From the formula   (\ref{jacobian}) one can see that the combination $\sqrt{g} T \bar \xi^0_0$  is the Jacobian $ det_3 \parallel \partial x/ \partial \beta \parallel $, and one can carry out the integration over $\beta$  on the right side.  So  finally:
\begin{eqnarray}
\label{erste}
& &\int d^3 \beta \sqrt{g} T \bar \xi^0_0 \bar \xi_b^m e^{\i px} A_m=\int  \frac{d^3k}{(2 \pi)^3 \sqrt{2k_0}} [ a_b(k) \delta(k-p)+   a^+_b(k) \delta(k+p)]= \nonumber \\
& &=\frac{1}{(2 \pi)^3 \sqrt{2p_0}}  [a_b(\vec k) + a^+_b(-\vec k)].
\end{eqnarray}

These  are four equations  and one needs some additional four to solve for the eight  unknown functions  $a_a,a_a^+$.

The canonical conjugate momentum is:
\begin{eqnarray}
\Pi_m=\frac{ \partial \mathcal{L}}{\partial( \partial_0 A^m) }.
\end{eqnarray}

In the formula for the Lagrangian  (\ref{fix}) the first term gives the following  contribution to the momenta:
\begin{eqnarray}
& &\Pi_m=- \sqrt{g} (\partial_0 A_m- \partial_m A_0) \nonumber \\
&= &- \sqrt{g} \int  \frac{d^3k}{(2 \pi)^3 \sqrt{2k_0}} \Bigl\{a_a(k) [\partial_0 \bar \xi^a_m- \i k_b \bar \xi^b_0   \bar \xi^a_m ] e^{-\i kx}+ a^+_a(k) [\partial_0 \bar \xi^a_m+ \i k_b \bar \xi^b_0 \bar \xi^a_m ] e^{\i kx} -      \nonumber \\
& &-  a_a(k) [\partial_m \bar \xi^a_0- \i k_b \bar \xi^b_m \bar \xi^a_0 ] e^{-\i kx}- a^+_a(k) [\partial_m \bar \xi^a_0+ \i k_b \bar \xi^b_m \bar \xi^a_0 ] e^{\i kx}      \Bigr\}  \nonumber \\
&=&\sqrt{g} \int  \frac{d^3k}{(2 \pi)^3 \sqrt{2k_0}} \Bigl\{ a_a(k) \i k_b [\bar \xi^b_0   \bar \xi^a_m- \bar \xi^b_m   \bar \xi^a_0  ] e^{-\i kx}-  \nonumber \\
& &-a^+_a(k) \i k_b [\bar \xi^b_0   \bar \xi^a_m- \bar \xi^b_m   \bar \xi^a_0  ] e^{\i kx} \Bigr\}
\end{eqnarray}

  Multiplying both sides with the expression $T \xi^m_c e^{\i px} $, for $t=0 $  (initial time) on the right side one gets  :
\begin{eqnarray}
& &a_a(k) k_b [\bar \xi^b_0   \bar \xi^a_m- \bar \xi^b_m   \bar \xi^a_0 ] \xi^m_c=
a_c(k) k_b \bar \xi^b_0 - a_a(k) k_c  \xi^a_0=  \nonumber \\
& &=[a_c(k) k_0-a_0(k) k_c] \bar \xi_0^0.
\end{eqnarray}

Here is used that for $t=0 \; (x_0=0) $ only the inverse Killing vector $ \bar  \xi^0$
 has a  non-vanishing zero component: $ \bar \xi_0^0 \not=0 $ and for the  other three vectors:  $ \bar \xi_i^0 =0 $ with $ i=1,2,3 $ (\ref{invkil}).

Integrating the last expression over the curved coordinates $\beta$ one gets:
\begin{eqnarray}
& &\int d^3 \beta T \xi^m_c e^{\i px} \Pi_m= \int  \frac{d^3k}{(2 \pi)^3 \sqrt{2k_0}} \int d^3 \beta \i \sqrt{g} T \bar \xi_0^0 \Bigl\{ [a_c(k) k_0-a_0(k) k_c] e^{-\i (k-p)x}-   \nonumber \\
& & [a^+_c(k) k_0-a^+_0(k) k_c] e^{\i (k+p)x}       \Bigr\}.
\end{eqnarray}

Remembering that $ \sqrt{g} T \bar \xi_0^0 $ is the Jacobian (\ref{jacobian}),with $T=1/(\cos t+ \cos \alpha)^2$, one can carry out the $x$ integration that leads to a delta functions.  At the last step one integrates over the momenta on the right side and finally:
\begin{eqnarray}
\label{zweite}
\int d^3 \beta T \xi^m_c e^{\i px} \Pi_m= \frac{1}{-\i\sqrt{2p_0} } \Bigl\{ a_c(p) p_0-a_0(p) p_c-a^+_c(-\vec p) p_0+a^+_0(- \vec p) \bar p_c   \Bigr\},
\end{eqnarray}

with $\bar p \equiv (p_0,-\vec p)$.

The second term in the Lagrangian (\ref{fix}) gives the following contribution:
\begin{eqnarray}
\label{pi'}
\Pi'_n= -\delta^0_n \frac{1}{T} \partial_k (\sqrt{g}T A^k )= -\delta^0_n \Bigl\{\frac{1}{T} \partial_k (\sqrt{g}T)g^{kl}A_l +  \sqrt{g} \partial_k  A_l g^{kl}  \Bigr\}.
\end{eqnarray}
For the  metric of $R \times S^3$ one has: $\partial_k g^{kl}=0$.
In the following  both sides will be multiplied with some unknown vector $f^0_c(\beta)$, with  $c=0,1,2,3. $ During the computations one can see which choice  is convenient for this vector (which allows us to carry  out the integration over the $\beta$ coordinates).  Integrating (\ref{pi'}) over $\beta$ one gets:

\begin{eqnarray}
&-&\int d^3 \beta \Pi_0'f^0_c e^{\i px}=\int  \frac{d^3k}{(2 \pi)^3  \sqrt{2k_0}}
\int d^3 \beta f^0_c g^{kl} \Bigl\{ \frac{1}{T} \partial_k (\sqrt{g}T) [a_a(k) \bar \xi^a_l e^{-\i (k-p)x}+ \nonumber \\
& &a^+_a(k) \bar \xi^a_l e^{\i (k+p)}]+\sqrt{g} \Bigl( a_a(k)[\partial_k \bar \xi^a_l- \i k_b \bar \xi^b_k \bar \xi^a_l] e^{-\i (k-p)x}+ a_a^+(k)[\partial_k \bar \xi^a_l+  \nonumber \\
& &+ \i k_b \bar \xi^b_k \bar \xi^a_l] e^{\i (k+p)x}  \Bigr)  \Bigr\}.
\end{eqnarray}

Using (\ref{fcomput}) one can see that on the space $R \times S^3$ the inverse conformal Killing-vectors fulfill the equality:
\begin{eqnarray}
\frac{1}{T}  \partial_k (\sqrt{g}T \bar \xi^a_l ) =g^{kl}  \left(\sqrt{g} \partial_k \bar \xi^a_l+  \frac{1}{T}  \partial_k (\sqrt{g}T)  \bar \xi^a_l     \right)=0.
\end{eqnarray}
So two terms remain on  the right side(containing the $\bar \xi^b_k \bar \xi^a_l$ products) and with (\ref{mtavari}) one gets:
\begin{eqnarray}
&-&\int d^3 \beta \Pi_0'f^0_c e^{\i px}=\int  \frac{d^3k}{(2 \pi)^3  \sqrt{2k_0}}
\int d^3 \beta  \sqrt{g} f^0_c \Bigl\{ -\i a_a(k) k_b \eta^{ab}T e^{-\i (k-p)x}+  \nonumber \\
& &+\i a^+_a(k) k_b \eta^{ab}T e^{\i (k+p)x}     \Bigr\}
\end{eqnarray}
One can see that  for $f^0_c= T \xi^0_c $  ($c=0,1,2,3  $) it is easy to take the integral on  the right side.  Actually  evaluating  this at  initial time $t=0 \, (x_0=0 )$ this vector has non-vanishing components  only for  $c=0$.   Again  we use the fact that the combination $ \sqrt{g}T \bar \xi^0_0$ is equal to the  three Jacobian of the transformation from the coordinates $(x_1,x_2,x_3)$ to the curved ones $(\beta_1,\beta_2,\beta_3)$.
\begin{eqnarray}
\label{dritte}
& &\int  \frac{d^3k}{(2 \pi)^3  \sqrt{2k_0}} \i \delta^0_c  \Bigl\{-a_a(k) k^a \delta(k-p)+a^+_a(k) k^a \delta(k+p)   \Bigr\}=  \nonumber \\
& &=\frac{\delta^0_c}{(-\i) \sqrt{2 p_0}} \left(-a_a(\vec p) p^a+ a^+_a(-\vec p)\bar p^a        \right).
\end{eqnarray}

The results (\ref{zweite}) and (\ref{dritte}) give:
\begin{eqnarray}
& &(-i)\sqrt{2p_0} \int d^3 \beta \Pi_m f^m_c e^{\i px} = \nonumber \\
& &=  a_c(p) p_0-a_0(p) p_c-a^+_c(-\vec p) p_0+a^+_0(- \vec p) \bar p_c                  +(a_a(\vec p) p^a- a^+_a(-\vec p)\bar p^a ) \delta^0_c.
\end{eqnarray}

Here  $f^m_c$ denotes the summarized four-vector that was multiplied in the  formulas (\ref{zweite})and (\ref{dritte}) to simplify the integration:
\begin{eqnarray}
f^m_c=
\left( \begin{array}{c}
T \xi^0_c        \\
T \xi^1_c       \\
T \xi^2_c      \\
T \xi^3_c
\end{array}  \right),
\end{eqnarray}

with $T$ from (\ref{tdef})  and the Killing-vectors from (\ref{veckill}).

Introducing the vectors:
\begin{eqnarray}
& &M_b(\vec p)=\sqrt{2 p_0} \int d^3 \beta \sqrt{g} T \bar \xi^0_0 \bar \xi_b^m e^{\i px} A_m,      \nonumber \\
& &N_b(\vec p)=\i \sqrt{2 p_0} \int d^3 \beta \Pi_m f^m_c e^{\i px},
\end{eqnarray}
one gets the following system of the eight algebraic equations for unknown functions $a_a(\vec k)$   and $a_a^+(-\vec k)$, with the subindex  $a=0,1,2,3 :$
\begin{eqnarray}
& &a_a+a^+_a=M_a   \nonumber \\
& &-a_0 p_0 +a_1p_1 +a_2p_2 +a_3p_3 +a^+_0p_0+a^+_1p_1+a^+_2p_2+a^+_3p_3=N_0 \nonumber \\
& &-a_1 p_0 +a_0p_1 +a^+_1p_0+ a^+_0p_1=N_1 \nonumber \\
& &-a_2 p_0 +a_0p_2 +a^+_2p_0+ a^+_0p_2=N_2 \nonumber \\
& &-a_3 p_0 +a_0p_3 +a^+_3p_0+ a^+_0p_3=N_3. \nonumber \\
\end{eqnarray}

The solution of this system for $x_0=0$ is:
\begin{eqnarray}
& &a_0(\vec p)=\frac{p_0 M_0+p_1M_1+p_2 M_2+p_3 M_3-N_0}{2 p_0} \nonumber \\
& &a_i(\vec p)=\frac{p_i M_0+p_0 M_i-N_i}{2p_0}  \nonumber \\
& &a_0^+(\vec p)=\frac{p_0 \bar M_0 +p_1 \bar M_1+p_2 \bar M_2 +p_3 \bar M_3 + \bar N_0}{2 p_0} \nonumber \\
& &a_i^+(\vec p)=\frac{p_i \bar M_0+p_0 \bar M_i+\bar N_i}{2p_0}.
\end{eqnarray}

where $\bar M(\vec p) \equiv M(-\vec p)$, $\bar N(\vec p) \equiv N(-\vec p)$
and  $i=1,2,3. $


\section{The Interactions with a scalar density field}
  One can introduce the  couplings with a vector field into the free scalar theory through a substitution in the scalar Lagrangian:
\begin{eqnarray}
\label{agaug}
p_m  \rightarrow p_m+\i  A_m,
\end{eqnarray}
where $A_m$ is a  gauge field. The conformally invariant Lagrangian of the charged  scalar field:
\begin{eqnarray}
\mathcal{L}_{\rm{scalar}}=g^{1/4} g^{mn} \partial_m \phi^{\ast} \partial_n \phi-\frac{1}{6} \hat R  \mid \phi \mid ^2
\end{eqnarray}
where $\hat R $ is a Ricci scalar corresponding  to the metric $ g^{1/4} g^{mn}$ and the $\phi's$ are  scalar densities  of  weight $1$. Together with the  free vector field  Lagrangian (\ref{lagmax}), and  using (\ref{agaug}), one gets:
\begin{eqnarray}
\label{totlag}
\mathcal{L}&=&\mathcal{L}_{\rm{maxwell}}+g^{1/4} g^{mn} (\partial_m-\i A_m)\phi^{\ast} (\partial_n+ \i A_n)\phi-\frac{1}{6} \hat R \mid \phi \mid ^2 \nonumber  \\
& &=\mathcal{L}_{rm{maxwell}}+\mathcal{L}_{\rm{scalar}}+ \i g^{1/4} g^{mn} A_m(\phi \partial_n \phi^{\ast} -\phi^{\ast} \partial_n \phi)+  \nonumber  \\
& & g^{1/4} g^{mn} A_m A_n \mid \phi \mid ^2.
\end{eqnarray}
 This Lagrangian describes the interaction of the scalar with an vector field.

  The last term in this expression is invariant under general coordinate transformations, because it is a product of the scalar $g^{mn} A_m A_n $ and the scalar density $g^{1/4}  \mid \phi \mid ^2$ of weight  $4$. As the metric tensor just enters through the conformally invariant combination: $g^{1/4} g^{mn}$, this term is invariant under conformal transformation of the fields $A_m$
\begin{eqnarray}
\delta_c ( g^{1/4} A^2 \mid \phi \mid ^2 )=\partial_n(\xi^n g^{1/4} A^2 \mid \phi \mid ^2  ).
\end{eqnarray}
 The third term is conformally invariant because $ \phi \partial_n \phi^{\ast} -\phi^{\ast} \partial_n \phi $ transforms as a vector density of weight $2$:
\begin{eqnarray}
& &\delta (\phi \partial_n \phi^{\ast} -\phi^{\ast} \partial_n \phi)=(\xi^l \partial_l \phi+\frac{1}{4} \partial_l \xi^l\phi ) \phi^{\ast}+\phi  (\xi^l \partial_l \phi^{\ast}+\frac{1}{4} \partial_l \xi^l\phi^{\ast} )- \nonumber \\
& &-(\xi^l \partial_l \phi^{\ast}+\frac{1}{4} \partial_l \xi^l\phi^{\ast} ) \phi-\phi^{\ast}  (\xi^l \partial_l \phi+\frac{1}{4} \partial_l \xi^l\phi )\nonumber \\
& &=\xi^l \partial_l(\phi \partial_n \phi^{\ast} -\phi^{\ast} \partial_n \phi)+\partial_n \xi^l (\phi \partial_l \phi^{\ast} -\phi^{\ast} \partial_l \phi)+\frac{1}{2} \partial_l \xi^l ( \phi \partial_n \phi^{\ast} -\phi^{\ast} \partial_n \phi)
\end{eqnarray}

 Hence  the whole third term in the Lagrangian is a scalar density of weight $4$.
 The equations of motion now are:
\begin{eqnarray}
\sqrt{g}D_m F^{mn}+\i g^{1/4}g^{mn}(\phi \partial_n \phi^{\ast} -\phi^{\ast} \partial_n \phi)+2 g^{1/4}g^{mn} A_m \mid \phi \mid ^2=0,
\end{eqnarray}
 and
\begin{eqnarray}
 \partial_m \left(g^{1/4}g^{mn}(\partial_n+\i A_n) \phi  \right)+\i g^{1/4}g^{mn}A_n (\partial_n+\i A_n) \phi+\frac{1}{6} \hat R \phi=0.
\end{eqnarray}

For the field $\phi $ one has:
\begin{eqnarray}
& &\phi=\int \frac{d^3 k}{(2 \pi)^3 \sqrt{2 k_0}} \left( b(k) w+ c^+(k) w^{\ast}\right)  \nonumber \\
& &\phi^+= \int \frac{d^3 k}{(2 \pi)^3 \sqrt{2 k_0}} \left(b^+(k)  w^{\ast}+ c(k) w\right)
\end{eqnarray}

 In the case of the scalar density the eigenfunctions of the momentum operator $w $ are:
\begin{eqnarray}
\label{scal}
w=g^{1/8} \sqrt{T} e^{-\i kx }
\end{eqnarray}
with $T$ from (\ref{tdef}) and $x^m$ from (\ref{xcoord}).   $b^+(k)$ creates  particles  with  momentum $k_m$ and $c^+(k)$ creates the corresponding antiparticles. For convenience we call the factor in front of the exponent $f$:

\begin{eqnarray}
\label{fff}
f \equiv g^{1/8} \sqrt{T}.
\end{eqnarray}

  In order to find  the vertex factor in the Feynman diagrams one must consider the interaction terms in the Lagrangian (\ref{totlag}).
The Fourier transformed of the fields $A_m(x)$ and  $\phi(x)$ have the form:
\begin{eqnarray}
\label{four}
& &A_m(x)=\int  d^4 p \hspace{0.2cm} \bar \xi^a_m e^{\i px }\tilde A_a(p)  \nonumber \\
& &\phi(x)=\int d^4 p\hspace{0.2cm}  g^{1/8} \sqrt{T} e^{\i px }\tilde \phi(p)
\end{eqnarray}

In the third term the derivatives $\partial_n \phi$ and $\partial_n \phi^{\ast}$ contain also derivatives of $f$. Next, this part will be investigated:
\begin{eqnarray}
& & \phi \partial_n \phi^{\ast}-\phi^{\ast} \partial_n \phi \rightarrow  \nonumber \\
& &\rightarrow \int \int  \frac{d^3 k}{(2 \pi)^3 \sqrt{2 k_0}} \frac{d^3 p}{(2 \pi)^3 \sqrt{2 p_0}} ( b(p) f e^{-\i px} (\partial_n f) b^+(k) e^{\i kx}+ \nonumber \\
& &+b(k) f e^{-\i px} (\partial_n f) c(k) e^{-\i kx}+ c^+(p) f e^{\i px} (\partial_n f) b^+(k)  e^{\i kx}+  \nonumber \\
& &+ c^+(p) f e^{\i p x} (\partial_n f) c(k) e^{- \i kx}-b^+(p) f e^{\i px} (\partial_n f) b(k) e^{-\i kx}- \nonumber \\
& &-b^+(k) f e^{\i px} (\partial_n f) c^+(k) e^{\i kx}- c(p) f e^{-\i px} (\partial_n f) b(k)  e^{-\i kx}-  \nonumber \\
& &-c(p) f e^{-\i px} (\partial_n f) c^+(k)  e^{\i kx} ).
\end{eqnarray}

  Exchanging in the last four terms the $k's $ and $p's$ there remain only four terms:
\begin{eqnarray}
& &\int \int  \frac{d^3 k}{(2 \pi)^3 \sqrt{2 k_0}} \frac{d^3 p}{(2 \pi)^3 \sqrt{2 p_0}}  [b(p),b^+(k)]f (\partial_n f) e^{\i (k-p)x}+[c^+(p),c(k)]f (\partial_n f) e^{-\i (k-p)x} \nonumber \\
& &= \left([b(p),b^+(k)]-[c(p),c^+(k)] \right) f (\partial_n f) e^{\i (k-p)x}=0.
\end{eqnarray}
Where in the last term $k$ and $p$ have been exchanged again. Furthermore it has been used that the $b$ and $c$ operators fulfill the same commutation relations:
\begin{eqnarray}
& &[b(p),b^+(k)]=(2 \pi)^3 \delta(p-k),  \nonumber \\
& &[c(p),c^+(k)]=(2 \pi)^3 \delta(p-k).
\end{eqnarray}

So the terms with derivatives of the functions $f$ ( \ref{fff}) vanish.

The last computation has shown that the vertex contribution from  the first part of the  third term  is:
\begin{eqnarray}
& & \i \int d^4 \beta \hspace{0.2cm} g^{1/4} g^{mn} A_m \phi \partial_n \phi^{\ast}  \nonumber \\
& &=\i \int d^4 \beta  \int \frac{d^4p}{(2 \pi)^4}   \frac{d^4k_1}{(2 \pi)^4}   \frac{d^4k_2}{(2 \pi)^4}   g^{1/4} g^{mn} \bar \xi^a_m  g^{1/4} T e^{\i (p+k_1)x}  \nonumber \\
& & (\partial_n e^{-\i k_2x })  \tilde A_a(p) \tilde \phi(k_1)  \tilde \phi^{\ast}(k_2)  \nonumber \\
& & \nonumber \\
& &=-\i \int d^4 \beta  \int \frac{d^4p}{(2 \pi)^4}   \frac{d^4k_1}{(2 \pi)^4}  \frac{d^4k_2}{(2 \pi)^4}   g^{mn}  g^{1/2} T \bar \xi^a_m  \i( k_2)_b \bar \xi^b_n  \nonumber \\
& & \times e^{\i (p+k_1-k_2)x} \tilde A_a(p) \tilde \phi(k_1)  \tilde \phi^{\ast}(k_2)  \nonumber \\
& & \nonumber \\
& &= \int \frac{d^4p}{(2 \pi)^4}   \frac{d^4k_1}{(2 \pi)^4}   \frac{d^4k_2}{(2 \pi)^4}\eta^{ab}(k_2)_b e^{\i (p+k_1-k_2)x}  \tilde A_a(p) \tilde \phi(k_1)  \tilde \phi^{\ast}(k_2)  \nonumber \\
& &= \int \frac{d^4p}{(2 \pi)^4}   \frac{d^4k_1}{(2 \pi)^4}   \frac{d^4k_2}{(2 \pi)^4}\eta^{ab}(k_2)_b (2 \pi)^4 \delta^4 (p+k_1-k_2) \nonumber \\
& &\times\tilde A_a(p) \tilde \phi(k_1)  \tilde \phi^{\ast}(k_2).
\end{eqnarray}

Then for the whole third term one has:
\begin{eqnarray}
& &\i \int d^4 \beta \hspace{0.2cm} g^{1/4} g^{mn} A_m( \phi \partial_n \phi^{\ast}-  \phi^{\ast}  \partial_n \phi)  \nonumber \\
& &= \int \frac{d^4p}{(2 \pi)^4}   \frac{d^4k_1}{(2 \pi)^4}   \frac{d^4k_2}{(2 \pi)^4}\eta^{ab}\left( (k_1)_b+(k_2)_b \right) (2 \pi)^4 \delta^4 (p+k_1-k_2) \nonumber \\
& &\times\tilde A_a(p) \tilde \phi(k_1)  \tilde \phi^{\ast}(k_2),
\end{eqnarray}
that gives the factor: $ ( (k_1)^a +(k_2)^a ) (2 \pi)^4 \delta^4 (p+k_1-k_2) $


From the fourth term in the Lagrangian one gets for the vertex factor:
\begin{eqnarray}
& &\i  \int d^4 \beta \hspace{0.2cm}  g^{1/4} g^{mn}\phi \phi^{\ast} A_m A_n  \nonumber \\
& &=\i \int d^4 \beta  \int \frac{d^4p_1}{(2 \pi)^4}   \frac{d^4p_2}{(2 \pi)^4}  \frac{d^4 k_1}{(2 \pi)^4}  \frac{d^4 k_2}{(2 \pi)^4} g^{1/2} Tg^{mn}  \bar \xi^a_m \bar \xi^b_n  \nonumber \\
& & \times   e^{-\i (p_1-p_2+k_1+k_2)x}  \tilde \phi(p_1)   \tilde \phi(p_2) \tilde A_a(k_1) \tilde A_b(k_2)  \nonumber \\
& & \nonumber \\
& &=\i \int \frac{d^4p_1}{(2 \pi)^4}   \frac{d^4p_2}{(2 \pi)^4}  \frac{d^4 k_1}{(2 \pi)^4}  \frac{d^4 k_2}{(2 \pi)^4} (2 \pi)^4 \delta^4(p_1-p_2+k_1+k_2) \nonumber \\
& & \times \tilde \phi(p_1)   \tilde \phi(p_2) \tilde A_a(k_1) \tilde A_b(k_2),
\end{eqnarray}
that gives the factor:   \quad    $ (2 \pi)^4 \delta^4(p_1-p_2+k_1+k_2)$    \quad



\newpage


\section{The energy spectrum}
Using  the formulas  (\ref{concurrent}) and (\ref{deltal}) one can write  the conserved current corresponding to the momentum operators $P_f$:
\begin{eqnarray}
\label{conc}
j^n_f&=& \delta A_m \frac{\partial \mathcal{L}}{\partial \partial_n A_m}-\xi^n_f \mathcal{L} \nonumber \\
&=& \delta A_m  \left[- \sqrt{g}F^{mn}- \frac{1}{T}g ^{nm} \partial_k(\sqrt{g}T g^{kl}A_l)\right]+  \nonumber \\
& &+\xi^n_f \left( \frac{1}{4}F_{mn}F^{mn}+\frac{1}{2\sqrt{g}T^2}[\partial_m(\sqrt{g}TA^m)]^2  \right).
\end{eqnarray}

Next   the offdiagonal terms will be computed, so in the mode expansion  for the   $A_m$'s (\ref{AA_m})   only the part $a_c u^c_m $ is considered.  For convenience it will be  used: $\tilde d k \equiv d^3 k/(2 \pi)^3 \sqrt{2 k_0}$. Using (\ref{eigvalue})  one gets for the first term in the last expression:
\begin{eqnarray*}
(W_1)^n_f=\int \tilde d k \int \tilde d p  \hspace{0.2cm} (-\i )k_f a_c(k) \bar \xi^c_m (-\sqrt{g}) g^{kn} g^{lm}\i a_a(p) p_b[\bar \xi^a_k \bar \xi^b_l-\bar \xi^a_l\bar \xi^b_k] e^{-\i(k+p)x}
\end{eqnarray*}
with (\ref{mtavari}) one further gets:
\begin{eqnarray}
(W_1)^n_f &=&-\int \tilde d k \int \tilde d p \hspace{0.2cm} k_f a_c(k)  \sqrt{g} g^{kn} a_a(p) p_b T[\eta^{bc} \bar \xi^a_k -\eta^{ac} \bar \xi^b_k]  e^{-\i(k+p)x} \nonumber \\
&=&-\int \tilde d k \int \tilde d p  \hspace{0.2cm} k_f   \sqrt{g} g^{kn} a_a(p) T \eta^{ab} \bar \xi^c_k [p_b a_c(k)-p_c a_b(k)]  e^{-\i(k+p)x}.
\end{eqnarray}

For the second term in the formula (\ref{conc}) one has:
\begin{eqnarray*}
(W_2)^n_f&=&\int \tilde d k \int \tilde d p  \hspace{0.2cm} \i k_f a_c(k) \bar \xi^c_m \frac{1}{T}  \Bigl[ \partial_k(\sqrt{g}T g^{kl}) a_a(p) \bar \xi^a_l e^{-\i(k+p)x}+\nonumber \\
& &+\sqrt{g}T g^{kl}  a_a(p) (\partial_k \bar \xi^a_l- \i p_b  \bar \xi^a_l  \bar \xi^b_k )   e^{-\i(k+p)x} \Bigr].
\end{eqnarray*}

One can compute that  the inverse Killing-vectors fulfill the following equality:
\begin{eqnarray}
\label{xi}
g^{kl} \partial_k (\sqrt{g}T \bar \xi^a_l )=0.
\end{eqnarray}
Using this, one can see that the first two terms in $W_2$ vanish  and that  only the last term remains:
\begin{eqnarray}
(W_2)^n_f=\int \tilde d k \int \tilde d p  \hspace{0.2cm} p_b k_f a_c(k) a_a(p) \bar \xi^c_k  g^{kn} \sqrt{g} T \eta^{ab} e^{-\i(k+p)x}.
\end{eqnarray}

This term cancels with the  first term of  $W_1$ and:
\begin{eqnarray}
(W_1)^n_f+(W_2)^n_f=\int \tilde d k \int \tilde d p  \hspace{0.2cm} k_f T a_a(p) a_b(k)  \eta^{ab} p_c\bar \xi^c_k \sqrt{g} g^{kn}  e^{-\i(k+p)x}.
\end{eqnarray}
From (\ref{mtavari}) one has:
\begin{eqnarray}
\bar \xi^d_k g^{kn} \eta_{ed} \eta^{ec}=T \xi^n_e \eta^{ec},
\end{eqnarray}
and finally:
\begin{eqnarray}
(W_1)^n_f+(W_2)^n_f=\int \tilde d k \int \tilde d p  \hspace{0.2cm}  k_f a_a(p) a_b(k) T^2 \xi^n_e \eta^{ec} \eta^{ab} p_c \sqrt{g}  e^{-\i(k+p)x},
\end{eqnarray}

The third term in (\ref{conc}) is:
\begin{eqnarray}
(W_3)^n_f&=&-\frac{1}{4}\xi^n_f \sqrt{g} g^{mk} g^{nl} \int \tilde d k \int \tilde d p  \hspace{0.2cm}  a_a(k)  k_b a_c(p) p_d (\bar \xi^a_m \bar \xi^b_n-\bar \xi^a_n\bar \xi^b_m) (\bar \xi^c_k \bar \xi^d_l-\bar \xi^c_l\bar \xi^d_k)  e^{-\i(k+p)x} \nonumber \\
&=&-\frac{1}{4}\xi^n_f  \sqrt{g}\int \tilde d k \int \tilde d p   a_a(k)  k_b a_c(p) p_d 2T^2 (\eta^{ac}\eta^{bd}- \eta^{ad}\eta^{bc}) e^{-\i(k+p)x}
\end{eqnarray}
and for the fourth term, from (\ref{conc}) and using once more the equality (\ref{xi}), there  remains:
\begin{eqnarray}
(W_4)^n_f=-\xi^n_f \frac{1}{2} \sqrt{g}T^2 \int \tilde d k \int \tilde d p  \hspace{0.2cm} \eta^{ab}a_a(p) p_b  \eta^{cd}a_c(k) k_d  e^{-\i(k+p)x}.
\end{eqnarray}

Adding up all four terms one gets:
\begin{eqnarray}
\label{jami}
(j_{\rm{offdiag1}})^n_f&=&(W_1)^n_f+(W_2)^n_f+(W_3)^n_f+(W_4)^n_f \nonumber \\
&=&-\frac{1}{2} \sqrt{g}T^2 \int \tilde dk \int \tilde dp  \hspace{0.2cm} e^{-\i(k+p)x} \eta^{ab}\eta^{cd}\Bigl[\xi^n_f a_a(p)p_b a_c(k)k_d \nonumber \\
&+&\xi^n_f a_a(p)p_c a_b(k) k_d  -\xi^n_fa_a(p) p_a a_c(k) k_b-2 \xi^n_da_a(p)p_f  a_b(k)k_c  \Bigr].
\end{eqnarray}
The subindex $1$ in $ j_{\rm{offdiag1}} $ means that it contains  only the annihilation operators.

One gets the same result as (\ref{jami}) for the part $ a_c^+ u_m^{c\ast} $ from the formula (\ref{AA_m}), but  the annihilation operators $a$ must be replaced by the creation operators $a^+$. The corresponing current is denoted as $j_{\rm{offdiag2}}$. So the part of the Hamiltonian corresponding to the offdiagonal terms  for  initial time $t=0 \, (x_0=0)$ is:
\begin{eqnarray}
H_1&=& \int d^3 \beta ( j_{\rm{offdiag}})^0_0 = \int d^3 \beta ( j_{\rm{offdiag1}})^0_0+( j_{\rm{offdiag2}})^0_0 \nonumber \\
&=&- \frac{1}{2}\int \frac{d^3 k}{(2 \pi)^3 2 k_0} \Bigl[ \eta^{ab} \eta^{cd}(a_a(-k)a_c(k) \bar k_b k_d +a_a(-k)a_b(k)  \bar k_c k_d- \nonumber \\
&-&a_a(-k)a_c(k) \bar k_d k_b )- 2 \eta^{ab} k_0^2 a_a(-k) a_b(k) \Bigr] +[a \to a^+].
\end{eqnarray}

Here is used that $\sqrt{g}T^2 \xi^0_0$ is the  determinant of the Jacobian $ \parallel \frac{\partial \beta}{\partial x} \parallel_3 $ and hence:
\begin{eqnarray}
\int d^3 \beta \sqrt{g}T^2 \xi^0_0 e^{-\i(k+p)x}=(2 \pi )^3 \delta(k+p).
\end{eqnarray}

The second and the fourth terms cancel:
\begin{eqnarray}
H_1= - \int \frac{d^3 k}{(2 \pi)^3 4 k_0} \eta^{ab} \eta^{cd} \left[a_a(-k)a_c(k) \bar k_b k_d-a_a(-k)a_c(k) \bar k_d k_b \right] +[a \to a^+],
\end{eqnarray}
one can give a simpler form to this expression:
\begin{eqnarray}
H_1&=& - \int \frac{d^3 k}{(2 \pi)^3  2 k_0} \eta^{ab} \eta^{cd} a_a(-k) a_c(k) k_0 [\delta^0_b k_d-\delta^0_d k_b ] +[a \to a^+] \nonumber \\
&=& -\int \frac{d^3 k}{2 (2 \pi)^3 } \eta^{ab} [a_0(-k)a_b(k)k_a- a_0(k)a_b(-k)k_a ]= +[a \to a^+] \nonumber \\
&=&\int \frac{d^3 k}{2 (2 \pi)^3} \delta^{ij} a_0(-k) a_j(k) k_i +[a \to a^+]
\end{eqnarray}
with $i,j=1,2,3$.

For the diagonal terms, containing   $a_a(k) a^+_b(p)$, one has an expression  similar to (\ref{jami}), but with  reversed signs and where the exponential function has a different power:
\begin{eqnarray}
\label{jami1}
& &(j_{\rm{diag1}})^n_f=\frac{1}{2} \sqrt{g}T^2 \int \tilde dk \int \tilde dp  \hspace{0.2cm} e^{-\i(k-p)x} \eta^{ab} \eta^{cd} \Bigl[\xi^n_f a_a^+(p)p_b a_c(k)k_d \nonumber \\
&+&\xi^n_f a_b(k)k_c a_a^+(p) p_d  -\xi^n_f a_c(k) k_b a_a^+(p) p_d-2 \xi^n_d a_b(k) k_f  a_a^+(p) p_c  \Bigr].
\end{eqnarray}

The remaining diagonal terms, containing  $a_a^+(k) a_b(p)$, give the same result as the last expression, so together the diagonal terms give (\ref{jami1}) without the factor $\frac{1}{2}$. In the Hamiltonian one finally  has for the time $x_0=0$:
\begin{eqnarray}
H_2&=&\int d^3 \beta 2 ( j_{\rm{diag1}})^0_0=\int \frac{d^3 k}{(2 \pi)^3 2 k_0} \Bigl[ \eta^{ab} \eta^{cd} \Bigl(a_a^+(k)a_c(k) k_b k_d +a_a^+(k)a_b(k)  k_c k_d- \nonumber \\
&-&a_a^+(k)a_c(k)  k_b k_d \Bigr)- 2 \eta^{ab} k_0^2 a_a^+(k) a_b(k) \Bigr] \nonumber \\
&=&-\int \frac{d^3 k}{(2 \pi)^3 } k_0 \eta^{ab} a_a^+(k) a_b(k).
\end{eqnarray}
 Here it was used again that:
\begin{eqnarray}
\int d^3 \beta \sqrt{g}T^2 \xi^0_0 e^{-\i(k-p)x}=(2 \pi )^3 \delta^3(k-p).
\end{eqnarray}
and the fact that only the Killing vector $\xi^m_0$ has nonvanishing zero component at time $x_0=0$ and for other three $\xi^0_i=0$ by $i=1,2,3.$ at this time.

The total Hamiltonian  is the sum of the diagonal and the offdiagonal parts and for the original time :
\begin{eqnarray}
\label{H1}
H&=&H_1+H_2=-\int \frac{d^3 k}{(2 \pi)^3 } k_0 \eta^{ab} a_a^+(k) a_b(k)+\nonumber \\
& &+\frac{1}{4k_0}\eta^{ab} \eta^{cd} \left[a_a(-k)a_c(k) \bar k_b k_d-a_a(-k)a_c(k) \bar k_d k_b \right]+\nonumber \\
& &+\frac{1}{4k_0} \eta^{ab} \eta^{cd}  \left[a_a^+(-k)a^+_c(k) \bar k_b k_d-a^+_a(-k)a^+_c(k) \bar k_d  k_b \right].
\end{eqnarray}

We will come back to this formula at the end of the next section.

\section{The Physical States}

In formula (\ref{H1}) not all creation and annihilation operators correspond to  physical states.  One introduces an  hermitian operator, the BRS operator: $s=s^+$.  This operator is nilpotent: $s^2=0$.

 One can decompose the creation operator, in the parts in the direction of the light-like momentum $k$,  in the direction $\bar k^a=(k^0,-k^1,-k^2,-k^3)$, and  in two directions $n^i,i=1,2 $ (orthogonal to those and one-another ).
\begin{eqnarray}
a^+_m(\vec k)=\sum_{\nu=k,\bar k,1,2} \epsilon_m^\nu a_\nu^+(\vec k).
\end{eqnarray}

Here $\epsilon_m^\nu$ is a polarization vector.  One possible choice is:
\begin{eqnarray}
\epsilon_a^\nu=\left(\frac{k_a}{\sqrt{2} k^0 },\frac{\bar k_a}{\sqrt{2} k^0},n^1_a,n^2_a   \right).
\end{eqnarray}

The  inverse vector is:
\begin{eqnarray}
\label{inve}
\bar \epsilon^a_\nu =\left(\frac{\bar k^a}{\sqrt{2} k^0 },\frac{- k^a}{\sqrt{2} k^0},(n^1)^a,(n^2)^a   \right),
\end{eqnarray}
because it is fulfilled:
\begin {eqnarray}
\epsilon^\nu \bar\epsilon^{\nu'}=\eta^{ab}\epsilon_a^\nu \bar \epsilon_b^{\nu'}=
\left( \begin{array}{cccc}
1  &    0  &    0  &    0    \\
0  &    1  &   0  &    0  \\
0  &     0  &    1 & 0 \\
0  & 0  &0  & 1
\end{array}  \right).
\end {eqnarray}

The scalar product of  two polarization vectors is:
\begin {eqnarray}
\epsilon^\nu \epsilon^{\nu'}=\eta^{ab}\epsilon_a^\nu \epsilon_b^{\nu'}=
\left( \begin{array}{cccc}
0  &    1  &    0  &    0    \\
1  &    0  &   0  &    0  \\
0  &     0  &    -1 & 0 \\
0  & 0  &0  & -1
\end{array}  \right).
\end {eqnarray}

The physical states have a gauge freedom:
\begin {eqnarray}
A_m'=A_m+\partial_m C,
\end{eqnarray}
where the field $C$ has to satisfy the equation:
\begin {eqnarray}
\label{c}
\partial_m(\sqrt{g}T g^{mn} \partial_m C  )=0
\end{eqnarray}

Using the result of the  formula (\ref{meore}) and the condition $k^2=0$ one can see that:
\begin{eqnarray}
\partial_m C=k_a \xi^a_m e^{- \i k x}.
\end{eqnarray}

So the general solution of the equation (\ref{c}) is the superposition:
\begin{eqnarray}
\partial_m C=\int \tilde d  k \left(c k_a \xi^a_m e^{- \i k x}+ c^+ k_a \xi^a_m e^{\i k x}  \right),
\end{eqnarray}
where $c$  and $c^+$ are the annihilation and  creation operators of  the field $C$.  Finally:
\begin{eqnarray}
C=\int \tilde d k \left(c  e^{- \i k x}+ c^+ e^{\i k x} \right).
\end{eqnarray}

The physical states   $\mathcal{N}$  are defined as  equivalence classes  of the  states that vanishes under the action of the BRS operator:
\begin{eqnarray}
\mathcal{N}=\{\vert \psi \rangle  :s \vert \psi \rangle =0, \vert \psi \rangle  mod  \quad  s \vert \Lambda  \rangle , \forall  \quad \vert \Lambda \rangle \}
\end{eqnarray}.
So the states  $\vert \psi \rangle $  and   $\vert \psi \rangle + s \vert \Lambda  \rangle $  are equivalent because of the nilpotency.

 One can add to the Lagrangian a term of the form: $ \i s X(\phi,\partial \phi, \dots)$, the total Lagrangian will remain invariant under the BRS transformations: $s \mathcal{L}_{\rm{total}}=0$.   Here as fields $\phi$ are understood the real bosonic vector field $A_m$, the  auxiliary fields $B$,  the real fermionic ghost  $C$ and antighost  fields $\bar C$.  One defines the action of the BRS operator on this fields as follows:
\begin{eqnarray}
& &s \bar C(x)=\i B(x),  \qquad s B(x)=0 , \nonumber \\
& &s A_m(x)=\partial_m C(x), \qquad s C(x)=0.
\end{eqnarray}

This fields have  ghost numbers:
\begin{eqnarray}
gh(\bar C)=-1 , \quad gh(B)=0, \quad gh(A_m)=0, \quad gh(C)=1.
\end{eqnarray}
The Lagrangian must have ghost number $0$ and so  $gh(X)=-1$. A possible choice for it is:
\begin{eqnarray}
X=\bar C \left(-\frac{1}{2}B+ \frac{4}{g^{1/4} T} \partial_m(g^{mn} T \sqrt{g} A_n)  \right).
\end{eqnarray}
Here $T$ is  always:  $T=(\cos t +\cos \alpha)^{-2}$.  Acting on this function with the BRS operator one gets:
\begin{eqnarray}
\label{sx}
& &\i s \left[\bar C \left(-\frac{1}{2}B+ \frac{1}{2 g^{1/4} T} \partial_m(g^{mn} T \sqrt{g} A_n)  \right) \right]=  \nonumber \\
& &=-\frac{B}{2} \left(-B+ \frac{1}{g^{1/4} T} \partial_m(g^{mn} T \sqrt{g} A_n)  \right)+ \frac{\i }{2 g^{1/4} T} \bar C \partial_m(g^{mn} T \sqrt{g} \partial_n C)  \nonumber \\
& &= \frac{1}{2} \left(B- \frac{1}{g^{1/4} T} \partial_m(g^{mn} T \sqrt{g} A_n)   \right)^2- \frac{1}{2 \sqrt{g} T^2} [\partial_m(g^{mn} T \sqrt{g} A_n)]^2+ \nonumber \\
& &+\frac{\i }{2 g^{1/4} T} \bar C \partial_m(g^{mn} T \sqrt{g} \partial_n C).
\end{eqnarray}

From the equation of motion for the field $B$ it follows that:
\begin{eqnarray}
\label{B}
B&=&\frac{1}{g^{1/4} T} \partial_m(g^{mn} T \sqrt{g} A_n) \nonumber \\
&= &\frac{1}{g^{1/4} T} \partial_m \left( g^{mn} T \sqrt{g} \int d^3 \tilde k
 (a^+_a \bar \xi^a_n e^{\i kx}+a_a \bar \xi^a_n e^{-\i kx}) \right) \nonumber \\
&=&\i \int d^3 \tilde k \frac{1}{g^{1/4} T} g^{mn} T \sqrt{g}  \xi^a_n \xi^b_m
 (a^+_a \bar \xi^a_n e^{\i kx}-a_a \bar \xi^a_n e^{-\i kx}) \nonumber \\
&=& \i g^{1/4} T \int d^3 \tilde k \eta^{ab} k_b  (a^+_a  e^{\i kx}-a_a  e^{-\i kx}).
\end{eqnarray}

For the field $ C$ one gets from the equation (\ref{sx}):
\begin{eqnarray}
\frac{ \delta \mathcal{L}}{\delta \bar C}=\frac{\i }{ g^{1/4} T}  \partial_m(g^{mn} T \sqrt{g} \partial_n C)=0,
\end{eqnarray}
and the solution of this  equation,  as was already seen, is:
\begin{eqnarray}
C=\int \tilde d k \left(c  e^{- \i k x}+ c^+ e^{\i k x} \right).
\end{eqnarray}

The variation of the Lagrangian with respect to   the field  $ C$ is:
\begin{eqnarray}
\delta  \mathcal{L}&=&\frac{\i }{ g^{1/4} T}  \bar C \partial_m(g^{mn} T \sqrt{g} \partial_n  \delta C)= -\partial_m\left( \frac{\i }{ g^{1/4} T}  \bar C \right)g^{mn} T \sqrt{g} \partial_n  \delta C \nonumber \\
&=&\i \partial_n \left(g^{mn} T \sqrt{g}  \partial_m  \frac{1 }{ g^{1/4} T}  \bar C  \right) \delta \bar C
\end{eqnarray}
Here has been used that  boundary terms vanish  under the integral.  The solution of the  equation
\begin{eqnarray}
\partial_n \left(g^{mn} T \sqrt{g}  \partial_m  \frac{1 }{ g^{1/4} T}  \bar C  \right)=0
\end{eqnarray}
has the form:
\begin{eqnarray}
\label{barc}
\bar C=g^{1/4} T \int \tilde d k \left(\bar c  e^{- \i k x}+ \bar c^+ e^{\i k x} \right).
\end{eqnarray}
where $c(k)$ and $\bar c^+(k)$ are the fermionic creation and annihilation operators.

Using the definition of action of the BRS operator on the field $\bar C$ : $s \bar C= \i B$ and the formulas (\ref{B})  and (\ref{barc}) one gets:
\begin{eqnarray}
s \bar c^+=- \eta^{ab} k_b a_a^+=-\epsilon^\nu_a \epsilon^{\nu'}_b \eta^{ab} a_{\nu} k_{\nu'}=-k_0 a_{\bar k}^+.
\end{eqnarray}

Hence for the one-particle states one has:
\begin{eqnarray}
\label{sbarc}
s \bar c^+ \vert 0  \rangle =-k_0 a_{\bar k}^+\vert 0  \rangle.
\end{eqnarray}

From the other definition $s A_m=\partial_m C$ one has:
\begin{eqnarray}
s a_m^+=k_m c^+.
\end{eqnarray}
 Decomposing the creation operators into the  four, earlier introduced,  orthogonal directions  and multiplying both sides  with the inverse polarization vector one gets:
\begin{eqnarray}
s a^+_\nu =\bar \epsilon^m_\nu k_m c^+,
\end{eqnarray}
 and finally using (\ref{inve}):
\begin{eqnarray}
\label{sa}
s a^+_k  \vert 0  \rangle =\sqrt{2} k_0 c^+ \vert 0  \rangle.
\end{eqnarray}

The formulas (\ref{sbarc}) and (\ref{sa}) show that $ a^+_k  \vert 0  \rangle $ and   $\bar c^+ \vert 0  \rangle $  are not invariant under the action of the BRS operator and so do not  correspond to  physical states.  \\
 The states  $ a_{\bar k}^+\vert 0  \rangle $  and $c^+ \vert 0  \rangle $  are equivalent to $0$. \\
 The two transverse creation operators $a_1^+$ and $a_2^+$  generate the  physical states.

  From  formula (\ref{acomutat}) one gets:
\begin{eqnarray}
 [a_\nu(k), a^+_{\nu'}(k')]=\bar \epsilon_\nu^a  \bar \epsilon_{\nu'}^b (2 \pi )^3 \eta_{ab}  \delta^3(k-k').
\end{eqnarray}

Using (\ref{inve}) one gets for the physical degrees of  freedom:
\begin{eqnarray}
& &[a_1(k), a^+_1(k')]=-(2 \pi )^3   \delta^3(k-k'), \nonumber \\
& &[a_2(k), a^+_2(k')]=-(2 \pi )^3   \delta^3(k-k').
\end{eqnarray}

In the formula for the Hamiltonian (\ref{H1}) one can see that the second and the third line correspond to the unphysical degrees of freedom.  Decomposing the operators in the second term of this expression one obtains:
\begin{eqnarray}
& &\frac{1}{4k_0}\eta^{ab} \eta^{cd} a_a(-k)a_c(k) \bar k_b k_d= \nonumber \\
& &=\frac{1}{4k_0} \eta^{ab} \eta^{cd} \epsilon^\nu_a \epsilon^{\nu'}_b   a_\nu(-k)\bar k_{\nu'}  \epsilon^\mu_c   \epsilon^{\mu'}_d  a_\mu(k)k_{\mu'} \nonumber \\
& &=k_0 a_k(-k) a_{\bar k}(k)
\end{eqnarray}
 and due to the fact that  $a_{\bar k}(k)$ are unphysical: $  a_{\bar k}(k) \vert phys \rangle =0$, this term gives no contribution to the energy-eigenvalues for physical states.  An analogous   computation for the last three terms in (\ref{H1}) show the same result, they are all unphysical. So the physical part of the  Hamiltonian is:
\begin{eqnarray}
H_{phys}=-\int \frac{d^3 k}{(2 \pi)^3 } k_0 \eta^{ab} a_a^+(k) a_b(k).
\end{eqnarray}
This result is analogous to the flat case.

\newpage

\chapter{Summary}
 The curved manifold  $R \times S^3$ is locally conformal to the Minkowski space.
The vector field is constructed on  $R \times S^3$.  The free field  Lagrangian on the curved space is invariant
 under  conformal transformations of the dynamical fields  $A_m(x)$.  The gauge-fixing term is not conformally  invariant,
  but it is invariant under  Poincar\'e transformations of the fields $A_m(x)$.  As the breaking of conformal invariance
 has occurred in the unphysical part of the Lagrangian, it would be interesting to investigate if the physical subspace
still maintains conformal symmetry. Propagator quantization is carried  out.  The energy-spectrum of the physical subspace
is analogous to the spectrum of flat quantum field theory.

 The interaction terms with the charged scalar density field are  also  conformally invariant.  It is shown that this terms
 introduce  no new vertex diagrams, but just the same that are known from scalar QED in Minkowski space.  The Lagrangian for the
free scalar density does not break conformal symmetry \cite{el}.

At the borders of the patch representing $R^4$ the local flat
coordinates are singular.  This makes it difficult to relate
fields that live on different patches.  Physically it should be
possible for fields to propagate from one patch to another, but
because of fixed points (points where the vector fields vanish) of
the Killing fields corresponding to the momentum operators in the
conformal algebra, this does not seem to be possible.  This leads
to the conclusion that the momentum operators of the conformal
  algebra are probably not a  suitable choice for the physical  momentum operators on $ R \times S^3$.  Another
possible choice for the operator corresponding to the energy would
be the one creating time translations $L_{-1,0}$.

\newpage

\appendix
\chapter{The Lorentz generator}

As an example, the generator $L_{-1,3}$ will be computed.  For this  consider the transformations:
\begin{eqnarray}
\delta y^{-1}=y^3-ay^{-1} \\
\delta y^3=y^{-1}-ay^3 \\
\delta y^m=-ay^m
\end {eqnarray}
where $m=1,2,4,5$, and $a=y^{-1} y^3$.
From (\ref{embed}) we get:
\begin{eqnarray}
\delta y^{-1}=-\sin t \delta t= \sin \alpha \cos \theta-\sin \alpha \cos \theta \cos^2 t \label{dx0} \nonumber
\end {eqnarray}
\begin{eqnarray}
\label{dt}
\delta t= -\sin t  \sin \alpha \cos \theta,
\end {eqnarray}
\begin{eqnarray}
\delta y^4=-\sin \alpha \delta \alpha=-\cos t \sin \alpha \cos \theta \cos \alpha  \nonumber
\end {eqnarray}
\begin{eqnarray}
\label{dalp}
\delta \alpha=\cos t \cos \alpha \cos \theta,
\end {eqnarray}
\begin{eqnarray}
\delta y^3&=&\delta(\sin \alpha \cos \theta)=\cos t-\cos t \sin^2 \alpha \cos^2 \theta   \nonumber \\
& & \cos \alpha \cos \theta \delta \alpha  -\sin  \alpha \sin \theta \delta \theta=\cos t(1-\sin^2 \alpha \cos^2 \theta ), \\
& &   \nonumber \\
\delta y^1&=&\cos \alpha \sin \theta \cos \phi \delta \alpha+\sin \alpha \cos \theta \cos \phi \delta \theta-\sin \alpha \sin \theta \sin \phi \delta \phi \nonumber \\
&=&-\cos t \sin^2 \alpha \cos \theta \cos \phi,  \\
& &   \nonumber \\
\delta y^2&=&\cos \alpha \sin \theta \sin \phi \delta \alpha+\sin \alpha \cos \theta \sin \phi \delta \theta-\sin \alpha \sin \theta \cos \phi \delta \phi \nonumber \\
&=&-\cos t \sin^2 \alpha \cos \theta \sin \phi.
\end {eqnarray}
If we use (\ref{dt}) and (\ref{dalp}) then we get from the last three formulas:
\begin{eqnarray}
\label{dtheta}
\delta \theta=\cot \alpha  \cot \theta \Bigl( \cos t \cos \alpha \cos \theta-\frac{\cos t}{\cos \alpha \cos \theta} +\cos t \frac{\sin^2 \alpha}{\cos \alpha} \cos \theta   \Bigr),
\end {eqnarray}

\begin{eqnarray}
& &\delta \alpha+\tan \alpha \cot \theta \delta \theta-\tan \alpha \tan \phi \delta \phi =-\cos t \frac{\sin^2 \alpha}{\cos \alpha} \cos \theta, \nonumber  \\
& &          \nonumber \\
& &\delta \alpha+\tan \alpha \cot \theta \delta \theta+\tan \alpha \cot \phi \delta \phi =-\cos t \frac{\sin^2 \alpha}{\cos \alpha} \cos \theta. \nonumber  \\
\end {eqnarray}
From the last two equations we obtain:
\begin{eqnarray}
(\tan \phi+\cot \phi) \delta \phi&=&0 \nonumber  \\
\frac{\delta \phi}{\cos^2 \phi}&=&0 \nonumber  \\
\delta \phi &=& 0.
\end {eqnarray}
After some computation  equation (\ref{dtheta}) gives:
\begin{eqnarray}
\delta \theta=-\frac{\cos t \sin \theta }{\sin \alpha}.
\end {eqnarray}
Finally we have deduced the generator:
\begin{eqnarray}
L_{-1,3}=-\sin t \sin \alpha \cos \theta \partial_t+\cos t \cos \alpha\cos \theta \partial_{\alpha}-\frac{\cos t \sin \theta }{\sin \alpha}\partial_{\theta}.
\end {eqnarray}

\newpage

\chapter{Checking the commutation relation}

 Now the following commutation relation will be proved:
\begin{eqnarray}
[D,P_1]=-P_1.
\end{eqnarray}
According to  (\ref{gen}) this commutator is equal to:
\begin{eqnarray}
& &[-L_{-1,4},L_{-1,1}+ L_{1,4}]=         \nonumber  \\
&=& [\sin t \cos \alpha \partial_t+\cos t \sin \alpha \partial_\alpha,-\sin t \sin \alpha \sin \theta \cos \phi \partial_t+ \nonumber  \\
& & \cos t \cos \alpha \sin \theta \cos \phi \partial_{\alpha}+ \frac{\cos t \cos \theta\cos \phi}{\sin \alpha} \partial _{\theta}- \frac{\cos t \sin \phi}{\sin \alpha \sin \theta} \partial_{\phi} +\sin \theta \cos \phi \partial_{\alpha} \nonumber  \\
& &+ \cot \alpha \cos \theta \cos \phi \partial_{\theta}- \cot \alpha \frac{\sin \phi}{\sin \theta} \partial_{\phi} ]  \nonumber  \\
&=& -\sin^2 t \cos^2 t \sin \theta \cos \phi \partial_\alpha+\sin t \cos t \sin \alpha \cos \alpha \sin \theta \cos \phi\partial_t- \nonumber  \\
& &\frac{\sin^2 t \cos \alpha  \cos \theta  \cos \phi}{\sin \alpha}\partial _{\theta}  +\frac{\sin^2 t \sin \phi\cos \alpha}{\sin \alpha \sin \theta }\partial_{\phi} \nonumber  \\
& &+\sin t \sin \alpha \sin \theta \cos \phi \partial_t-\sin t\cos t \sin \alpha \cos \alpha \sin \theta \cos \phi \partial_t \nonumber  \\
& &-\sin^2 t  \sin^2 \alpha \sin \theta \cos \phi \partial_\alpha-\cos^2 t  \sin^2 \alpha \sin \theta \cos \phi \partial_\alpha- \nonumber  \\
& &\cos^2 t \cos^2 \alpha \sin \theta \cos \phi \partial_\alpha -\frac{\cos^2 t \sin \alpha  \cos \theta  \cos \phi }{\sin^2 \alpha}\partial _{\theta} \nonumber  \\
& &+\frac{\cos^2 t \sin \alpha \sin \phi }{\sin^2 \alpha \sin \theta} \partial_{\phi}- \cos t \cos \alpha \sin \theta \cos \phi \partial_{\alpha}-\frac{\cos t \sin \alpha \cos \theta \cos \phi }{\sin^2 \alpha} \partial_{\theta}+ \nonumber  \\
& &\frac{\cos t \sin \phi}{\sin \alpha \sin \theta}\partial_\phi  \nonumber  \\
&=&\sin t \sin \alpha \sin \theta \cos \phi \partial_t+(-1-\cos t \cos \alpha) \sin \theta  \cos \phi \partial_\alpha  \nonumber  \\
& &-\frac{\cos t+ \cos \alpha}{\sin \alpha}\cos \theta \cos \phi \partial_\theta+\frac{\cos t+ \cos \alpha}{\sin \alpha \sin \theta} \nonumber  \\
&=&-( L_{-1,1}+ L_{1,4})
\end {eqnarray}
 Thus the commutation relation is fulfilled.


\newpage
{\Large Selbst\"andigkeitserkl\"arung}
\vspace{1cm}

Hiermit versichere ich,die vorliegende Diplomarbeit selbstst\"andig angefertigt und keine anderen als die aufgef\"uhrten Hilfsmittel verwendet zu haben. \\



\begin{thebibliography}{99}
\bibitem{dr}
N. Dragon, Geometrie der Relativit\"atsteorie, http://www. itp. uni-hannover. de/ $\tilde{ }$ dragon 2001
\bibitem{dr1}
N. Dragon, BRS Symmetry and Cohomology, http://www. itp. uni-hannover. de/ $\tilde{ }$ dragon,1996

\bibitem{el}
Elizabeth Nuncio Quiroz, Diploma thesis  at the Institut f\"ur Theoretishe Physik, Hannover 2002


\bibitem{fe}
B. Felsager, Geometry, Particles and Fields, Springer,1998

\bibitem{ha}
S. W. Hawking and G. F. R. Ellis, The large scale structure of space-time,  Cambrige University Press, Cambrige,1986

\bibitem{se}
S. M. Paneitz  and I. E. Segal, Analysis in Space-Time Bundles. I General Considerations  and the scalar Bundle,Journal of Functional Analysis 47,78-142,(1982)

\bibitem{du}
B. A. Dubrovin,A. T. Fomenko and S. P.  Novikov, Modern Geometry-Methods and applications,Part 1, Springer-verlag,New York, 1992

\bibitem{la}
L.D.Landau, The Classical Theory of fields, Vol. 2, Addison-Wesley,Pergamon Press:London,1971

\end{thebibliography}
\end{document}